\documentclass[reprint,aps,prb,floatfix,superscriptaddress,longbibliography]{revtex4-2}
\usepackage[utf8]{inputenc}

\usepackage{amsmath,amssymb,braket}
\usepackage{bm}
\usepackage{graphicx}
\usepackage{booktabs}
\usepackage{enumitem}
\usepackage{bbm}
\usepackage{physics}

\usepackage{xcolor}
\usepackage[
    colorlinks=true,
    citecolor=blue,
    linkcolor=blue,
    urlcolor=blue
]{hyperref}

\newcommand{\JAB}{J_{AB}}
\newcommand{\JBB}{J_{BB}}

\newcommand{\mb}[1]{\mathbf{#1}}
\begin{document}

\title{Dynamical signatures of deconfined spinons in dimerized sawtooth chains}

%\title{From triplons to deconfined spinons in dimerized sawtooth chains}

%\title{Dynamical structure factor of the dimerized phase of the antiferromagnetic sawtooth chain}

\author{Nishan Ranabhat}
\email{uve4wq@virginia.edu}
\affiliation{Department of Physics, University of Virginia,
             Charlottesville, Virginia 22904, USA}

\author{Brandon B. Le}
\affiliation{Department of Physics, University of Virginia,
             Charlottesville, Virginia 22904, USA}
             
\author{Seung-Hun Lee}
\affiliation{Department of Physics, University of Virginia,
             Charlottesville, Virginia 22904, USA}

\author{Gia-Wei Chern}
\affiliation{Department of Physics, University of Virginia,
             Charlottesville, Virginia 22904, USA}

\date{\today}

\begin{abstract}
We study the ground-state properties and dynamical response of the Heisenberg model on the sawtooth chain in and away from the exactly solvable valence-bond-solid (VBS) point. We employ U(1)-symmetric density-matrix renormalization group (DMRG) and time-dependent variational principle (TDVP) methods to compute equilibrium diagnostics and the zero-temperature dynamical structure factor (DSF) $S^{zz}(q,\omega)$ across the dimerized phase, a valence-bond-ordered state in which the two symmetry-equivalent apex--base bonds of each triangle develop unequal spin correlations, probing the approach to the continuous lower phase boundary, the exact VBS point, and the approach to the first-order upper boundary. In every regime, the DSF is a broad continuum dominated by a bright band at its lower edge, with the intensity at the one-triplon energy $\omega \simeq J_{AB}$ suppressed. We identify the spectrum as a deconfined two-spinon continuum of kink and antikink domain walls between the two degenerate singlet coverings. Closed-form spinon dispersions fix the continuum edges and track the dominant band across the zone in all three regimes, while an explicit finite-separation two-kink calculation in a constrained Hilbert space reproduces the measured intensity distribution. Our results provide a microscopic picture of fractionalized excitations in the dimerized sawtooth chain and are relevant to the recently discovered Ti$^{3+}$ kagome fluorides, where strongly anisotropic exchange interactions can generate sawtooth-chain building blocks.
\end{abstract}

\maketitle

% =====================================================================
\section{Introduction}
\label{sec:intro}

Competing exchange interactions on geometrically frustrated lattices generate some of the richest collective behavior in condensed matter: fractionalized excitations, macroscopically degenerate ground-state manifolds, and exotic ordered and disordered phases. Quantum fluctuations are particularly strong in one dimension, where they can preclude magnetic order and lead to elementary excitations with fractional quantum numbers. The antiferromagnetic sawtooth chain, which is a line of corner-sharing triangles, combines these one-dimensional quantum effects with geometric frustration and provides a tractable setting in which dimerization and fractionalized excitations can be studied microscopically. This geometry is realized in real materials, including the natural mineral atacamite, Cu$_2$Cl(OH)$_3$~\cite{Heinze_2021}, high-spin molecular rings~\cite{Baniodeh_2018}, and, more recently, Ti$^{3+}$ kagome fluorides~\cite{Thennakoon_unpub}.

\begin{figure}[t]
  \centering
  \includegraphics[width=\columnwidth]{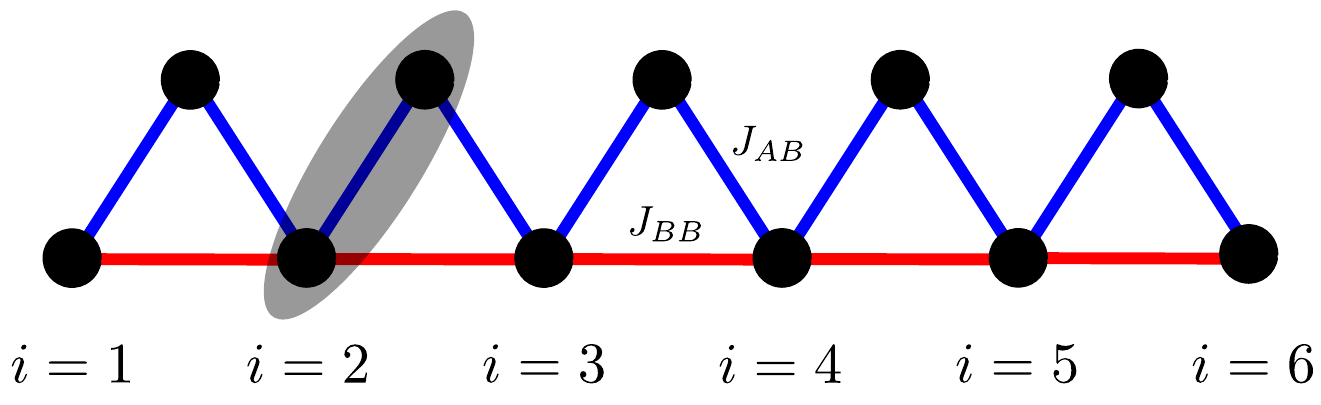}
\caption{Geometry of the sawtooth chain in Eq.~\eqref{eq:H}, with apical spins $\mathbf{S}^{A}_{i}$ and basal spins $\mathbf{S}^{B}_{i}$. The apex-base and base-base exchange interactions are denoted by $J_{AB}$ (blue) and $J_{BB}$ (red), respectively. The index $i$ labels the unit cell; the shaded region indicates one unit cell containing a basal and an apical spin.}
    \label{fig:geometry}
\end{figure}

The sawtooth chain (Fig.~\ref{fig:geometry}) has two spin-$1/2$ sites per unit cell: a basal/chain spin $\mathbf{S}^{B}_{i}$ and an apical spin $\mathbf{S}^{A}_{i}$, for $i=1,\ldots,L$ unit cells ($N=2L$ spins). With apex-base coupling $J_{AB}$ and base-base coupling $J_{BB}$, the Hamiltonian is
\begin{equation}
H = J_{BB}\sum_{i=1}^{L-1}\mathbf{S}^{B}_{i}\cdot\mathbf{S}^{B}_{i+1} + J_{AB}\sum_{i=1}^{L}\mathbf{S}^{A}_{i}\cdot\mathbf{S}^{B}_{i} + J_{AB}\sum_{i=1}^{L-1}\mathbf{S}^{A}_{i}\cdot\mathbf{S}^{B}_{i+1},
\label{eq:H}
\end{equation}
i.e., each apex couples to the two basal spins of its triangle, and consecutive basal spins couple along the chain. The single control parameter is the ratio $r\equiv J_{BB}/J_{AB}$ which can also be viewed as a frustration parameter. At $r=0$ the basal bonds vanish and Eq.~\eqref{eq:H} reduces to the uniform, unfrustrated $S=1/2$ Heisenberg chain of $N=2L$ spins with exchange $J_{AB}$, whereas at $r=1$ all three bonds of every triangle are equal and the frustration within each triangle is maximal. As $r$ varies at fixed antiferromagnetic sign, the ground state passes through three regimes~\cite{Blundell_2003,Jiang_2015,Rausch_2025}. At small $r$ ($r\lesssim0.49$), it is gapless and adiabatically connected to the $r=0$ Heisenberg chain, whose dynamics is characterized by the des Cloizeaux--Pearson two-spinon continuum~\cite{dCP_1962,Muller_1981}. At intermediate $r$ ($0.49\lesssim r\lesssim1.5$), the system enters a gapped dimerized phase: a valence-bond-ordered state in which the two symmetry-equivalent apex-base bonds of each triangle develop unequal spin correlations. This bond order is static, the two bond energies are fixed ground-state expectation values whose inequality breaks the mirror symmetry between the two bonds. Away from $r=1$, the ground state is a dressed superposition of many singlet configurations rather than a product of isolated dimers, yet the average apex-base bond contrast remains finite. The phase contains the exact valence-bond-solid (VBS) point $r=1$, where the bond contrast is maximal, and the ground state is a tensor product of apex-base singlets. At large $r$ ($r\gtrsim1.5$), it enters a gapless noncollinear phase through a first-order transition~\cite{Rausch_2025}. In this work, we study the dimerized phase, including both its equilibrium properties and its full dynamical response.

The Ti$^{3+}$ ($3d^1$, $S=1/2$) kagome fluorides~\cite{Goto_2016,Jiang_2020,Jiang_2023} have recently emerged as a family of frustrated magnets that interpolates between near-ideal kagome lattices and networks with strongly anisotropic exchange interactions. Density-functional theory calculations on these materials find strongly bond-dependent exchange interactions that, in the more distorted members, naturally generate linear and sawtooth ($\Delta$)-chain motifs~\cite{Jeschke_2019,Shirakami_2019}. Neutron scattering measurements on the modulated kagome antiferromagnet Cs$_8$RbK$_3$Ti$_{12}$F$_{48}$ revealed a gapless dispersive continuum indicative of fractionalized excitations~\cite{Thennakoon_2025}, while recent measurements on its sister compound Cs$_8$LiNa$_3$Ti$_{12}$F$_{48}$~\cite{Thennakoon_unpub,Bakshi_2026} further point to strongly anisotropic magnetic interactions. These developments motivate a microscopic understanding of the spin-$1/2$ sawtooth chain, not only as a paradigmatic frustrated one-dimensional system, but also as a building block relevant to this new class of Ti-based kagome magnets.

The VBS point ($r=1$) of the dimerized phase has been well studied~\cite{NakamuraKubo,SSWC,HaoTchernyshyov}: the ground state is an exact tensor product of apex-base singlets, and the elementary excitations are gapped kink and antikink spinons. A local spin excitation, however, carries $\Delta S=1$ and can also be viewed as creating an on-dimer triplet. This raises a basic question for the dynamical response: whether the spectral weight is governed by a coherent triplon excitation or by its fractionalized spin-$1/2$ constituents. Away from the solvable point, variational estimates of the antikink dispersion have been obtained~\cite{PaulGhosh}, but the dressed regimes $r\neq1$ remain largely unexplored, both in their ground-state properties and in their dynamical response.

The natural probe of the latter is the dynamical structure factor (DSF; see Sec.~\ref{sec:methods-dsf}), the momentum- and energy-resolved spin response measured directly by inelastic neutron scattering. A complete map of the zero-temperature $S^{zz}(q,\omega)$ across the dimerized phase, and with it the identification of the physical origin of its dominant low-energy excitations, remains unstudied. In particular, it remains unclear how the kink--antikink description at the exactly solvable VBS point evolves away from $r=1$, and whether fractionalized spinons continue to control the dynamical response throughout the dimerized phase. Addressing this question also traces how the dynamics evolves between the two distinct limits that bound the dimerized phase.

Methodologically, we employ a combination of tensor-network algorithms for ground-state and real-time dynamics~\cite{White_1992,Schollwock_2011,Haegeman_2016,Collura_2024}, variational probes~\cite{Feynman_1954,GMP_1986,Arovas1988,Sharma_2025}, and effective domain-wall analysis~\cite{HaoTchernyshyov,NakamuraKubo,SSWC}, akin to the recent study of partially dimerized $J_1$--$J_2$ chains~\cite{Sharma_2025}. Building on this toolkit, we compute and interpret the full zero-temperature longitudinal dynamical structure factor $S^{zz}(q,\omega)$ of the dimerized sawtooth chain at three representative points, $r=0.72$, $1.00$, and $1.38$, chosen to probe the approach to the lower continuous boundary, the exact VBS point, and the approach to the upper first-order boundary, respectively. We find that the dynamical response throughout the dimerized phase is dominated not by a coherent one-triplon mode, but by a broad two-spinon continuum formed by deconfined kink and antikink domain walls. Closed-form spinon dispersions determine the continuum boundaries and track its dominant low-energy spectral features, while an explicit finite-separation two-kink construction reproduces the distribution of spectral weight within the continuum.

The remainder of the paper is organized as follows. Section~\ref{sec:methods} describes the numerical and analytical methods. Section~\ref{sec:gs} presents the equilibrium ground-state characterization, while Secs.~\ref{sec:dsf}--\ref{sec:b1} present the dynamical structure factor and its two-spinon interpretation. Technical details are provided in Appendixes~\ref{app:tn}--\ref{app:twospinon}.

\section{Methods}
\label{sec:methods}

Our numerical approach combines three complementary methods built around a matrix-product-state (MPS) representation of the many-body wave function. We first use the density-matrix renormalization group (DMRG) to obtain accurate ground states and characterize their equilibrium properties. Starting from these ground states, we then employ the time-dependent variational principle (TDVP) to simulate the real-time evolution following a local spin excitation, from which the dynamical structure factor $S^{zz}(q,\omega)$ is obtained. We further use a numerical single-mode approximation (SMA), constructed directly from the DMRG ground state, as an independent variational probe of the excitation spectrum. These MPS-based methods are complemented by a semi-analytical finite-$D$ two-kink construction, anchored by brute-force exact diagonalization, which resolves the distribution of spectral weight within the dynamical continuum. Together, these methods provide complementary access to the ground-state properties, the full dynamical response, and the low-energy excitation spectrum of the sawtooth chain.

\subsection{Ground state from DMRG}
\label{sec:methods-dmrg}

All tensor-network calculations use $U(1)$-symmetric, block-sparse MPS tensors that explicitly conserve $S^z_{\rm tot}$. Ground states are obtained with the two-site density-matrix renormalization group (DMRG) algorithm~\cite{White_1992,Schollwock_2011} in the $S^z_{\rm tot}=0$ sector. We consider two system sizes with open boundary conditions (OBC): $L=64$ to study the equilibrium ground-state properties (see Sec.~\ref{sec:gs}), and $L=32$ to generate the ground states that enter the dynamical structure factor (DSF) protocol of Sec.~\ref{sec:methods-dsf}. The two-site update grows the bond dimension adaptively up to $\chi_{\rm DMRG}=128$, with SVD truncation cutoffs of $10^{-8}$--$10^{-10}$. The quality of the converged states is tested by the energy variance per site, $v=(\expval{H^2}-E_0^2)/N\JAB^2$, evaluated on the final MPS ($v=0$ for an exact eigenstate). For the $L=64$ states, $v=6.0\times10^{-11}$ for $r=0.72$ and $v=5.8\times10^{-9}$ for $r=1.38$, while for $r=1$ DMRG reproduces the exact valence-bond-solid ground state, $E_0=-\tfrac34L\JAB$, to machine precision ($v\lesssim10^{-14}$). For $L=32$, $v$ is marginally smaller for all three cases. Further details on the convergence and errors in DMRG are provided in Appendix~\ref{app:tn}.

\subsection{Dynamical structure factor protocol}
\label{sec:methods-dsf}

We compute the zero-temperature DSF, defined as the space-time Fourier transform of the dynamical spin-spin correlator,
\begin{equation}
  S^{zz}(q,\omega)=\int_{-\infty}^{\infty}\!\!dt\;e^{i\omega t}\,
  \frac1N\sum_{j,l}e^{-iq(r_j-r_l)}\,
  \langle S^z_j(t)\,S^z_l(0)\rangle,
  \label{eq:dsf}
\end{equation}
with $S^z_j(t)=e^{iHt}S^z_j e^{-iHt}$ and the expectation taken in the DMRG ground state $|\Psi_0\rangle$. The DSF is directly accessible in inelastic neutron scattering experiments~\cite{Thennakoon_2025}. Its real-time formulation also provides a natural route for its calculation within the MPS framework. Acting with a local spin operator on the ground state creates a localized excitation containing the many-body eigenstates that couple to this probe. The subsequent real-time evolution encodes their energies and propagation, which are resolved in momentum and frequency through Fourier transformation. We carry out this evolution using two-site TDVP, which projects the Schr\"odinger equation onto the MPS variational manifold while allowing the bond dimension to grow as entanglement develops during the time evolution.

With the ground state available, the protocol for calculating the DSF is as follows: a single $S^z$ operator is applied at the central cell $j_0=L/2$, separately on each sublattice site $\sigma\in\{B,A\}$, $|\phi^\sigma\rangle=S^z_{\sigma,j_0}|\Psi_0\rangle$. The real-time evolution of each source is then $|\phi^\sigma(t)\rangle=e^{-iHt}|\phi^\sigma\rangle$~\cite{Haegeman_2016,Collura_2024}. We use a maximum bond dimension $\chi_{\rm TDVP}=600$ and a time step $\Delta t=0.01\,\JAB^{-1}$, evolving up to the final time $t_{\rm final}=8\,\JAB^{-1}$. The truncation is monitored through the cumulative discarded Schmidt weight, which stays below $7.5\times10^{-4}$ in the least favorable channel. Further details on the errors incurred in the TDVP calculation are provided in Appendix~\ref{app:tn}. Measuring $S^z$ on every cell $j$ and sublattice $\sigma'$ against the evolved source gives the real-space Green's function
\begin{equation}
\begin{aligned}
  G^{\sigma'\sigma}(j,t)
  &=\langle\Psi_0|\,S^z_{\sigma',j}(t)\,S^z_{\sigma,j_0}(0)\,|\Psi_0\rangle\\
  &=\langle\Psi_0|e^{iHt}\,S^z_{\sigma',j}e^{-iHt}\,S^z_{\sigma,j_0}\,|\Psi_0\rangle\\
  &=e^{iE_0t}\langle\Psi_0|\,S^z_{\sigma',j}\,|\phi^\sigma(t)\rangle
\end{aligned}
  \label{eq:greens}
\end{equation}
in the four sublattice channels $(\sigma',\sigma)\in\{B,A\}^2$. Finally, we perform a double Fourier transform of $G^{\sigma'\sigma}(j,t)$ and sum over the sublattice contributions, $S^{zz}(q,\omega)=\tfrac1N\sum_{\sigma'\sigma}S^{\sigma'\sigma}(q,\omega)$, to evaluate Eq.~\eqref{eq:dsf}. Accounting for the relative positions of the two sublattices, with $\delta_B=0$ and $\delta_A=\tfrac12$, places the DSF in the extended Brillouin-zone convention $q\in[0,4\pi]$ used throughout.

\subsection{Numerical single-mode approximation}
\label{sec:methods-sma}

We complement the full numerical dynamical response calculated via TDVP with an independent variational estimate of the low-energy excitation spectrum. The single-mode approximation constructs a restricted variational space of excited states by acting on the ground state with a chosen set of local operators. Here we implement a multi-operator SMA~\cite{Feynman_1954,GMP_1986,Arovas1988}, in the matrix-valued form of Ref.~\cite{Sharma_2025}, directly on the MPS-based DMRG ground state $|\Psi_0\rangle$, retaining separate local spin operators on the two sublattices of the sawtooth chain.

The two local $S^z$ probes per cell, $\{\Omega^{B}_i=S^z_{B_i},\,\Omega^{A}_i=S^z_{A_i}\}$, span the $2L$-dimensional space of mutually nonorthogonal trial states $|\phi_i^{\sigma}\rangle=\Omega^\sigma_i|\Psi_0\rangle$. The variational spectrum follows from the single real-space generalized eigen-problem,
\begin{equation}
  \tilde H\,|\psi\rangle=\omega\,\mathcal O\,|\psi\rangle,
  \label{eq:sma-main}
\end{equation}
with overlap $\mathcal O^{\sigma\sigma'}_{ij}=\langle\Psi_0|(\Omega^\sigma_i)^\dagger\Omega^{\sigma'}_j|\Psi_0\rangle$ and $\tilde H^{\sigma\sigma'}_{ij}=\langle\Psi_0|(\Omega^\sigma_i)^\dagger(H-E_0)\Omega^{\sigma'}_j|\Psi_0\rangle$, built by sandwiching the Hamiltonian MPO between the corresponding MPS states. Since OBC break exact translation symmetry, momentum is assigned a posteriori: each eigenvector is projected onto the plane-wave combinations $|k,\sigma\rangle=L^{-1/2}\sum_j e^{-ikj}\Omega^\sigma_j|\Psi_0\rangle$ on the grid $k=2\pi m/L$ ($m=0,1,\ldots,L-1$) and labeled by the momentum of maximal weight. The lowest eigenvalue at each $k$ defines the variational SMA dispersion [black points in Fig.~\ref{fig:headline}(b),(d),(f)] and provides a variational estimate of the lowest excitation accessible within this local-operator subspace. At the exact VBS point, $r=1$, the SMA reproduces the flat triplon at $\omega=\JAB$. Regularization of the near-singular metric and filtering of the spurious $U(1)$ null mode are detailed in Appendix~\ref{app:sma}.

\subsection{Finite-$D$ two-kink construction and exact diagonalization}
\label{sec:methods-twokink}

To resolve how the spectral weight is distributed within the dynamical continuum, we diagonalize the effective two-spinon (kink-antikink) problem directly in real space. The pair state $\ket{i,d}$ [Fig.~\ref{fig:cartoon}(c)] is constructed explicitly for separations $0\leq d\leq D$, its exact overlap and Hamiltonian matrix elements are evaluated on sparse product states, and the resulting generalized eigen-problem $H(q)\psi=\omega\,\mathcal O(q)\psi$ is solved at each total momentum by canonical orthogonalization of the nonorthogonal metric. The DSF is obtained from the spectral weight of the $d=0$ source (Appendix~\ref{app:twospinon}). The truncation parameter $D$ is the maximum kink--antikink separation, with $D=0$ corresponding exactly to the coincident pair, or local triplon. Increasing $D$ therefore provides a direct way to follow how a locally created $S=1$ excitation changes as its two spin-$1/2$ constituents are allowed to separate.

The construction is anchored by exact diagonalization (ED) in the reduced Hilbert space. The closed-form spinon dispersions entering the two-kink analysis determine only energy differences, so the single additive constant $E_K$ is fixed by matching the bottom of the two-spinon continuum to the ED singlet--triplet gap: the gap is computed by sparse Lanczos diagonalization of periodic rings of $N=4$--$12$ unit cells and extrapolated linearly in $1/N$ to the production size $N=32$ (Appendix~\ref{app:twospinon}). ED further serves as an independent benchmark of the DMRG ground states at small system sizes.

\begin{figure*}[t]
  \centering
  \includegraphics[width=0.85\textwidth]{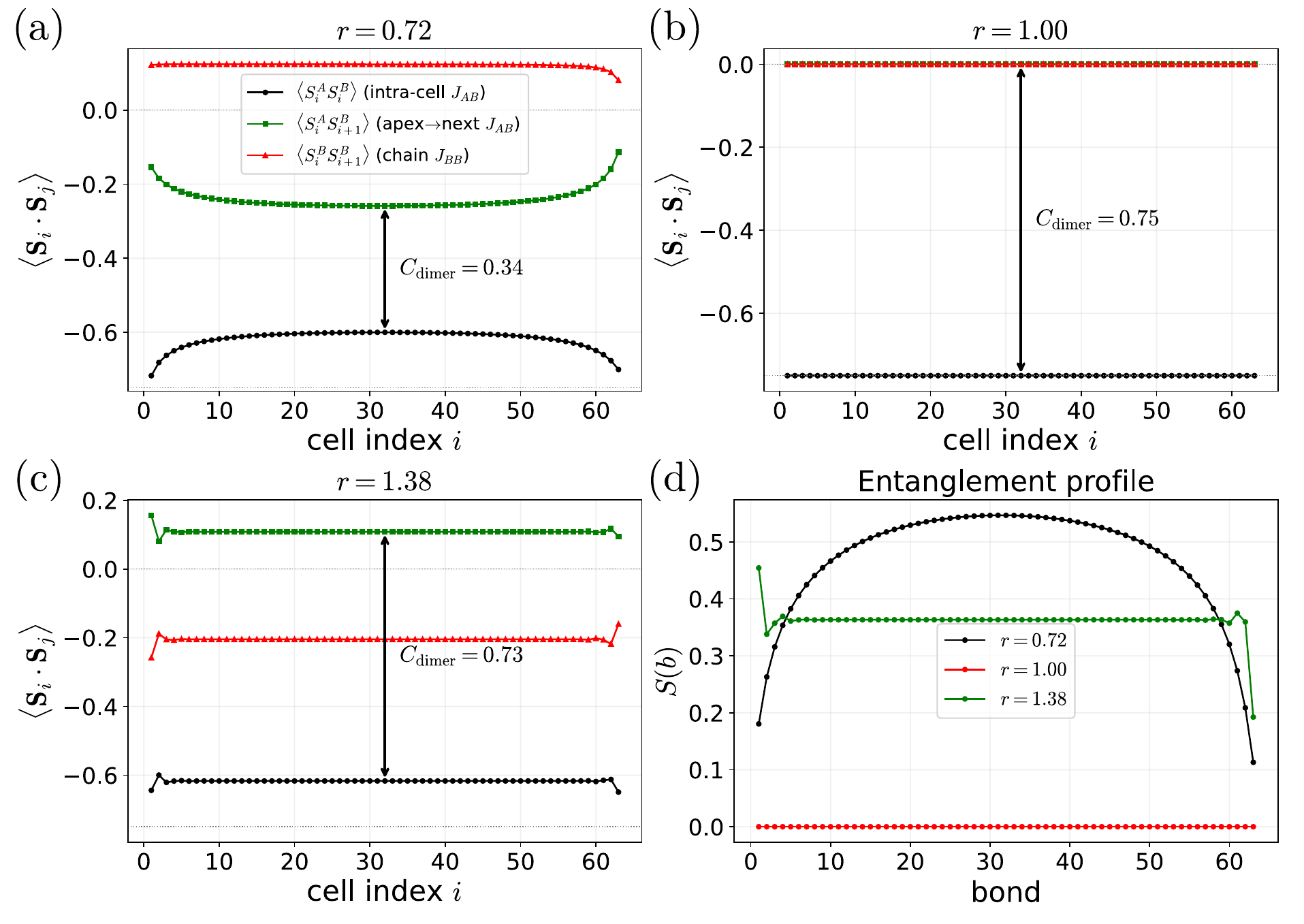}
  \caption{\label{fig:gs}%
    Ground-state characterization at $L=64$, $\chi_{\rm DMRG}=128$, from two
    different diagnostics of the dimerization. (a)-(c)~Bond-resolved
    ground-state energy $\langle\mb S_i\!\cdot\!\mb S_j\rangle$ for
    (a)~$r=0.72$, (b)~$r=1.000$, and (c)~$r=1.38$, decomposed into the
    intra-cell apex-base bond ($\langle\mb S^A_i\!\cdot\!\mb S^B_i\rangle$,
    $\JAB$; black), the apex-to-next-cell bond
    ($\langle\mb S^A_i\!\cdot\!\mb S^B_{i+1}\rangle$, $\JAB$; green), and the basal bond ($\langle\mb S^B_i\!\cdot\!\mb S^B_{i+1}\rangle$, $\JBB$; red); the dimer order parameter $C_{\rm dimer}$ is marked in each. (d)~Bipartite entanglement entropy $S(b)$ across each bond $b$ for the
    three regimes. Both diagnostics collapse to the exact valence-bond solid at
    $r=1$: perfect intra-cell singlets ($-\tfrac34$) with vanishing inter-cell
    bonds and $C_{\rm dimer}=\tfrac34$ [panel~(b)], and $S(b)=0$ on every
    inter-cell cut [panel~(d)]. The profiles are flat across the bulk (uniform
    dimerization; deviations only within a few cells of the open ends).}
\end{figure*}

% =====================================================================
\section{Results}
\label{sec:results}

\subsection{Ground-state characterization across the dimerized phase}
\label{sec:gs}

We study the ground state for an OBC system of size $L=64$ with DMRG at three
regimes within the gapped dimerized phase,
which occupies $0.49\lesssim r\lesssim 1.5$~\cite{Blundell_2003,Rausch_2025} and is bounded by two transitions of opposite character. Its lower boundary is a continuous transition into a gapless phase adiabatically connected to the uniform antiferromagnetic Heisenberg chain~\cite{Blundell_2003}, whose low-energy excitations are the des~Cloizeaux-Pearson two-spinon
continuum~\cite{dCP_1962, Muller_1981}; its upper boundary is a first-order transition
into the gapless noncollinear phase studied by ~\cite{Rausch_2025}. Our study bridges these two
limits, the Heisenberg chain below and the noncollinear phase above, through
the exactly solvable VBS at the center: $r=0.72$ probes the approach to the
continuous lower edge, $r=1.00$ the exact low entangled VBS, and $r=1.38$ the approach to the
first-order upper edge.

\paragraph{Dimer order and entanglement.}
Figure~\ref{fig:gs} characterizes the ground state using two complementary
diagnostics of the dimer order: the bond-resolved energy [panels
(a)-(c)] and the bipartite entanglement entropy at every bond [panel~(d)]. Panels (a)-(c)
resolve the energy onto the three inequivalent bonds in the sawtooth triangle. At the VBS point
[Fig.~\ref{fig:gs}(b)] the intra-cell apex-base bonds are exact singlets
($\langle\mb S\!\cdot\!\mb S\rangle=-\tfrac34$) while the apex-to-next-cell and
chain bonds carry zero energy: the ground state is a product of decoupled
apex-base singlets; consequently the dimer order $C_{\rm dimer} = \langle\mb S^A_i\!\cdot\!\mb S^B_{i+1}\rangle - \langle\mb S^A_i\!\cdot\!\mb S^B_i\rangle$, attains its
maximal value $\tfrac34$. The other two points behave
differently; at $r=1.38$ [Fig.~\ref{fig:gs}(c)] the strong dimerization survives
almost intact ($C_{\rm dimer}=0.73$), very close to the VBS value although it lies near the upper phase boundary
$r_{\mathrm{upper}}\!\approx\!1.5$. This is a direct signature of the
first-order phase transition between the dimer phase and the noncollinear phase: rather than slow
continuous decay to zero, the order strongly survives right up to the phase boundary and abruptly
drops to $\sim\!10^{-4}$ across it ~\cite{Rausch_2025}. At $r=0.72$ [Fig.~\ref{fig:gs}(a)] the bond contrast is already substantially reduced
($C_{\rm dimer}=0.34$) as the apex-to-next-cell bonds strengthen, the growing
antiferromagnetic correlations of the uniform zigzag ($\JAB$) chain that the
small-$r$ limit approaches, and the dimerization begins to
melt, the expected precursor of a continuous transition.
In every regime, the pattern is uniform through the bulk, confirming a homogeneously
dimerized ground state with open-boundary effects at the edges.

\begin{figure}[t]
  \centering
  \includegraphics[width=\columnwidth]{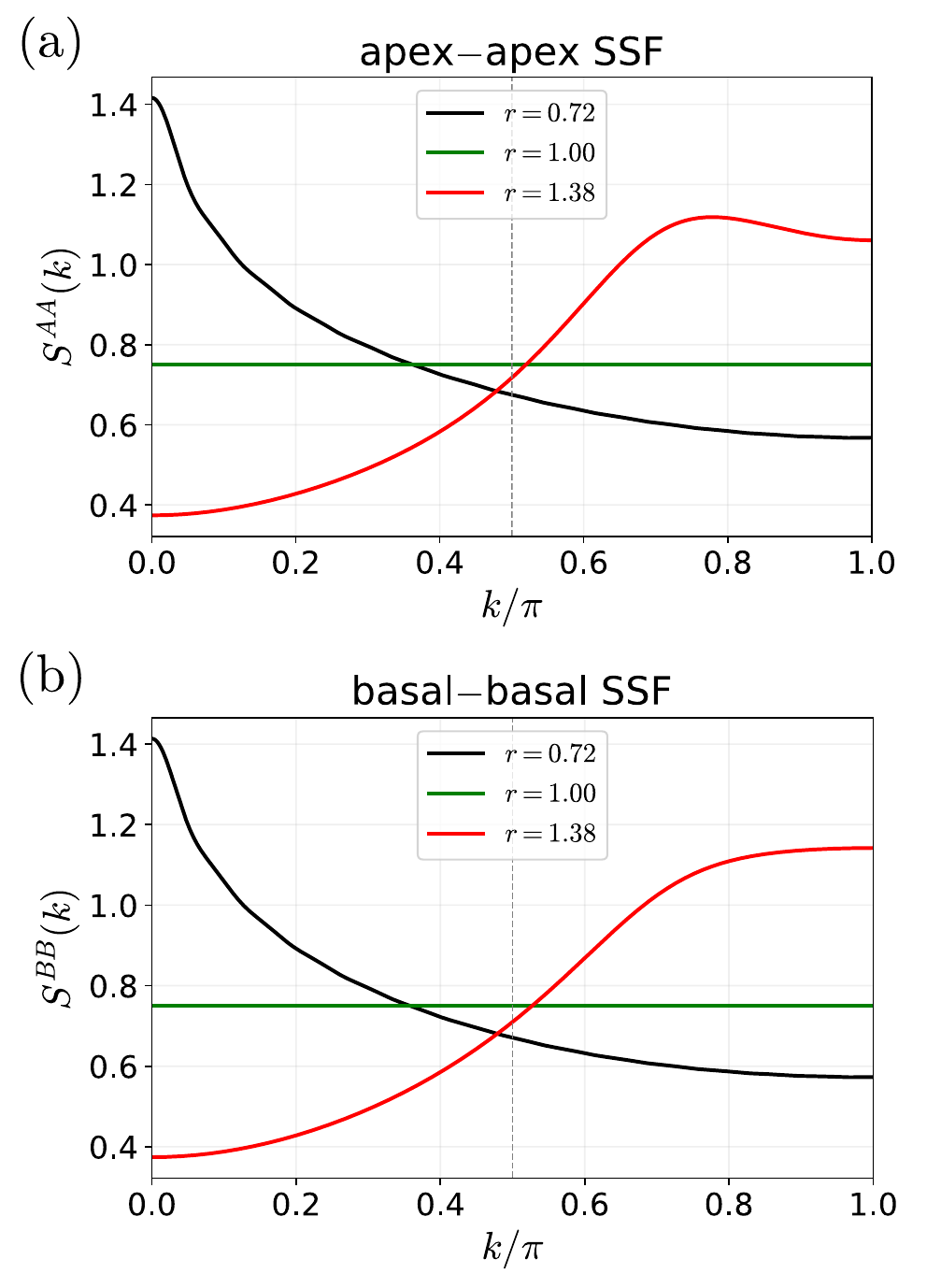}
  \caption{\label{fig:ssf}%
    Equilibrium static spin structure factor $S^{\sigma\sigma}(k)$
    [Eq.~\eqref{eq:ssf}] of the $L=64$ ground state, resolved by sublattice:
    (a)~apex-apex $S^{AA}(k)$ and (b)~basal-basal $S^{BB}(k)$, for $r=0.72$ (black), $1.000$ (green), and $1.38$
    (red). At the VBS point, the same-sublattice correlations are those of decoupled singlets and
    $S(k)$ is flat at $\tfrac34$. The dominant correlation wavevector migrates
    from $k=0$ for $r<1$ (proximity to the quasi-Heisenberg regime) toward
    $k=\pi$ for $r>1$, the two curves crossing near $k=\pi/2$ (dashed line), the wavevector at which the gap closes at the upper transition into the
    noncollinear phase~\cite{Rausch_2025}.}
\end{figure}

The entanglement entropy tells the same story from a
complementary angle. For a cut on bond $b$ that
partitions the chain into left and right blocks, the bipartite von Neumann entanglement entropy ~\cite{CalabreseCardy_2004} of the ground state is
\begin{equation}
  S(b) = -\Tr\!\big(\rho_L\ln\rho_L\big)
       = -\sum_{\alpha}\lambda_{\alpha}^2(b)\,\ln\lambda_{\alpha}^2(b),
  \label{eq:entropy}
\end{equation}
where $\rho_L=\Tr_R|\Psi_0\rangle\langle\Psi_0|$ is the reduced density matrix of
the left block and $\{\lambda_{\alpha}(b)\}$ are the Schmidt coefficients across the cut
(the MPS singular values at bond $b$, with $\sum_{\alpha}\lambda_{\alpha}^2=1$), so that $S(b)$
can be extracted directly off the DMRG MPS state. For a product state, $\lambda_{\alpha} = \delta_{1,\alpha}$; therefore, the entanglement entropy vanishes ($S(b)=0$) on every inter-cell bond. This is clearly observed for $r=1$ in Fig.~\ref{fig:gs}(d), confirming that the ground state is an exact tensor product
of the individual apex-base singlets, precisely the valence-bond solid. Panels [(a)-(c)] and (d) in Fig.~\ref{fig:gs} are therefore the dual signatures of the same fact: the singlets
are perfect and decoupled through their energies, and the absence of entanglement between them. Away from the dimer point, the two regimes carry finite entanglement with qualitatively different profiles
that distinguish the two phase boundaries. At $r=1.38$ the entropy is flat
across the bulk, saturating to a low, bond-independent plateau
($S\!\approx\!0.36$): the state is gapped and obeys the area
law~\cite{Hastings_2007,Eisert_2010}, consistent with
the first-order upper transition, on approach to which the gap drops discontinuously into the gapless noncollinear phase at the boundary
itself~\cite{Rausch_2025}). At $r=0.72$, by contrast, $S(b)$ shows a
pronounced dome shape peaking at the chain center, the finite-size hallmark of proximity
to the continuous lower transition, where the gap closes and correlations
become long-ranged, approaching the logarithmic (critical) profile of the uniform
antiferromagnetic chain~\cite{Vidal_2003,CalabreseCardy_2004}. The flat-versus-domed contrast thus reads
directly as first-order (gap-preserving) versus continuous (gap-closing) behavior
at the two edges of the dimerized phase.

\paragraph{Static structure factor.}
The same physics can be read in momentum space. The sublattice-resolved static
(equal-time) structure factor is the Fourier transform of the ground-state
spin-spin correlation function,
\begin{equation}
  S^{\sigma\sigma}(k)=\expval{\mathbf S^\sigma_0\!\cdot\!\mathbf S^\sigma_0}
  +2\sum_{d\ge1}\cos(kd)\,\expval{\mathbf S^\sigma_0\!\cdot\!\mathbf S^\sigma_{d}},
  \qquad \sigma\in\{A,B\},
  \label{eq:ssf}
\end{equation}
the position of its maximum is the dominant
correlation wavevector, its height and width measure the strength and range of the
correlations, and a flat $S^{\sigma\sigma}(k)$ signals the absence of any
inter-site correlation. Figure~\ref{fig:ssf} shows $S^{AA}(k)$ (apex) and
$S^{BB}(k)$ (basal). At the VBS point, the same sublattice correlations vanish for
every $d\ge1$, so $S^{\sigma\sigma}(k)=\expval{\mathbf S^\sigma_0\!\cdot\!\mathbf
S^\sigma_0}=\tfrac34$ is exactly flat, another momentum-space signature of the
decoupled-dimer VBS, alongside the bond energies [Fig.~\ref{fig:gs}(a)-(c)] and the
vanishing entanglement [Fig.~\ref{fig:gs}(d)]. Away from the dimer point, it
sits at $k=0$ for $r=0.72$ for both apex-apex and basal-basal spins, signifying ferromagnetic-like same sublattice
alignment, whereas for $r=1.38$ we observe two different behaviors. The basal
channel locks to $k=\pi$, signifying commensurate staggered antiferromagnetism, whereas the apical channel peaks at an incommensurate
wavevector, $k\approx0.78\pi$, already displaced from $\pi$ and moving toward $k=\pi/2$. This
apical shift is the equilibrium signature of the noncollinear order that sets in beyond the upper transition, whose ordering wavevector
approaches $k=\pi/2$~\cite{Rausch_2025}.

Taken together, these ground-state diagnostics establish that the three representative couplings probe distinct regimes within the same dimerized phase. The $r=0.72$ state already shows substantial weakening of the dimer order and enhanced entanglement as the continuous lower boundary is approached, whereas the $r=1.38$ state retains strong dimerization while developing incommensurate correlations characteristic of the nearby first-order boundary. With this ground-state evolution across the dimerized phase established, we now turn to the central question of this work: how these distinct regimes are reflected in the dynamical spin response.

\subsection{Dynamical structure factor}
\label{sec:dsf}

\begin{figure*}[t]
  \centering
  \includegraphics[width=0.85\textwidth]{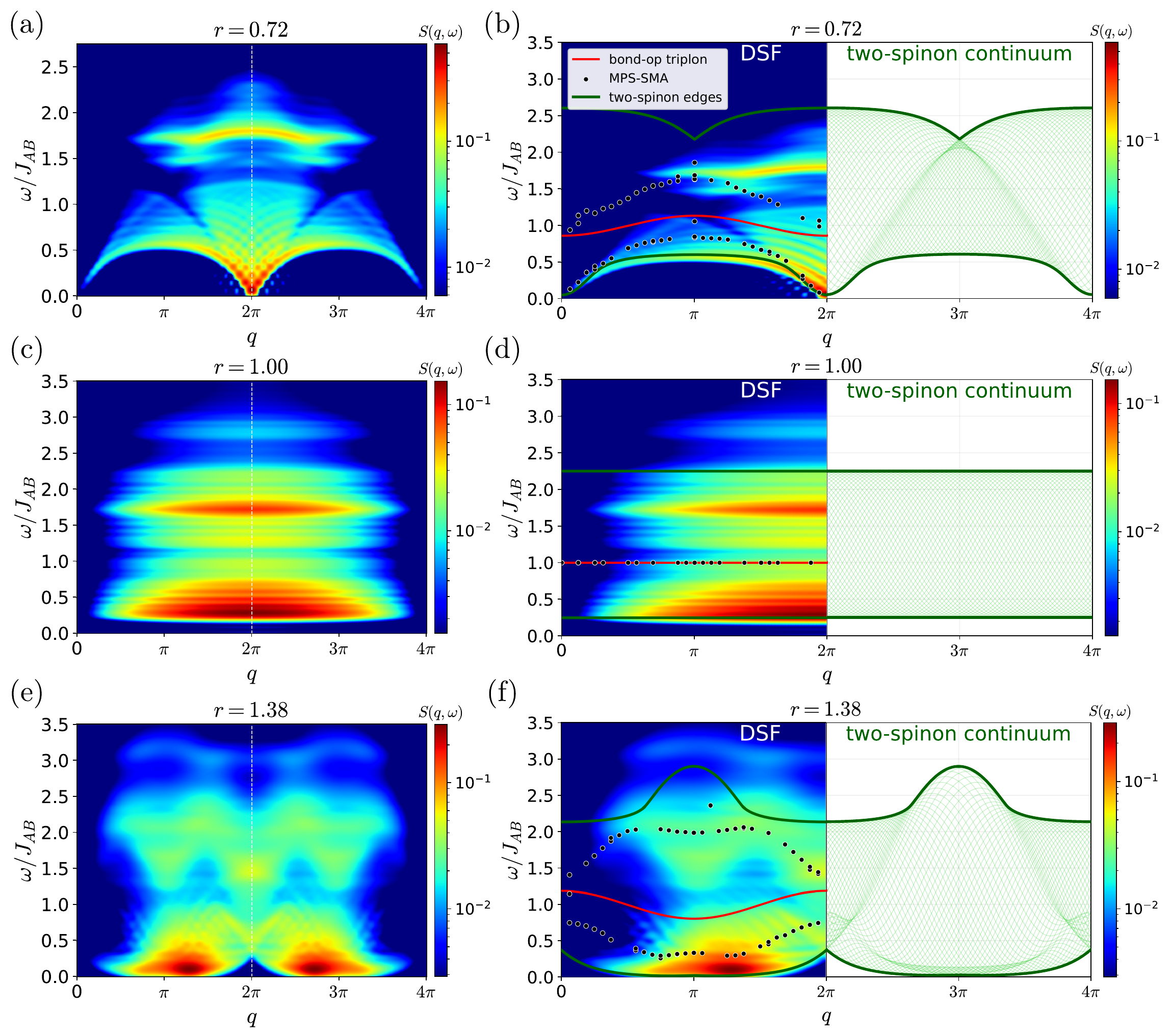}
  \caption{\label{fig:headline}%
    Dynamical structure factor $S^{zz}(q,\omega)$ of the dimerized
    sawtooth chain ($L=32$, $\chi_{\rm TDVP}=600$, $t_{\max}=8/\JAB$) across the three regimes: (a),(b)~$r=0.72$; (c),(d)~$r=1.000$ (exact VBS);
    (e),(f)~$r=1.38$. All panels are plotted in the physical (apex-offset,
    extended-zone) convention $q\in[0,4\pi]$; energies are in units of $\JAB$.
    Left column [(a),(c),(e)]: the measured DSF over the full extended zone. Right column [(b),(d),(f)]: interpretation of the same data; the left half ($q\in[0,2\pi]$) repeats the TDVP DSF with the analytic/semi-analytic overlays, and the right half ($q\in[2\pi,4\pi]$) shows the two-spinon continuum. Overlays: the two-spinon kink--antikink continuum [thick green = kinematic
    edges from the deconfined convolution Eq.~\eqref{eq:convol}; faint green =
    dense internal-momentum sampling], the bond-operator triplon
    $\omega_t(q)$ (red), and the numerical MPS-SMA modes (black
    points).}
\end{figure*}

Figure~\ref{fig:headline} summarizes the central numerical result of this work: the dynamical structure factor $S^{zz}(q,\omega)$ across the dimerized phase. The left column [(a),(c),(e)] presents the TDVP results over the full extended zone, while the right column [(b),(d),(f)] overlays the analytical and semi-analytical results developed below. Despite the substantial differences in the ground states at the three representative couplings, their dynamical responses share a common qualitative structure. The spectral weight forms a broad continuum dominated by a bright band near its lower edge. The remaining weight spreads into a fainter continuum above it, gathering into a weak secondary maximum in its upper part, with an intensity dip near $\omega\simeq\JAB$ in between. Qualitatively similar continua have been reported for partially dimerized $J_1$--$J_2$ chains~\cite{Sharma_2025}.

A natural starting point for understanding the excitation spectrum of a dimerized valence-bond state is the local $S=1$ triplet excitation of a singlet bond, or triplon. Within the self-consistent bond-operator treatment described in Appendix~\ref{app:bondop}, the corresponding one-triplon dispersion for the sawtooth chain is
\begin{equation}
\omega_t(q)=\JAB+\frac{\JBB-\JAB}{2}\cos q .
\end{equation}
At the exact VBS point $r=1$, this reduces to a flat mode at $\omega_t=\JAB$, while away from the VBS point the triplon becomes dispersive, with its minimum shifting from $q=0$ for $r<1$ to $q=\pi$ for $r>1$. This conventional triplon picture therefore provides a simple reference for the dynamical response throughout the dimerized phase. Strikingly, however, the TDVP spectra in Fig.~\ref{fig:headline} are not dominated by this mode, but instead by the broad continua described above.

As described in Sec.~\ref{sec:methods-dsf}, we plot the DSF in the physical (apex-offset) convention over the extended zone $q\in[0,4\pi]$. The apical-site offset $\delta_A=\tfrac12$ introduces a geometric form factor that makes $S^{zz}(q,\omega)$ asymmetric about $q=\pi$ and symmetric about $q=2\pi$, so that $[0,4\pi]$ is the natural window for comparison with neutron scattering. More remarkably, the spectral weight is suppressed in the vicinity of the expected one-triplon band (red curves in Fig.~\ref{fig:headline}), rather than concentrated around it. This suppression is particularly evident at the exact VBS point, where the flat triplon at $\omega=\JAB$ lies directly in the intensity dip separating the dominant lower band from the faint upper maximum. The persistence of this behavior throughout the dimerized phase suggests that the dominant dynamical excitations are not conventional local triplets. We first describe the numerical spectra across the three representative regimes and then develop their microscopic interpretation in terms of fractionalized kink and antikink excitations.

\paragraph{Exact VBS point ($r=1$).}

At the dimer point [Fig.~\ref{fig:headline}(c),(d)], the spectrum becomes a momentum-independent flat continuum spanning $\omega\simeq\tfrac14\JAB$ to $\tfrac94\JAB$. Its intensity is dominated by a flat band near the lower edge and a fainter secondary maximum near $\omega\simeq\tfrac74\JAB$, separated by an intensity dip near $\omega\simeq\JAB$. The latter is precisely the energy at which the one-triplon mode is expected.

The one-triplon mode, captured by the self-consistent bond-operator theory~\cite{SachdevBhatt} (Appendix~\ref{app:bondop}) and the MPS-SMA (Appendix~\ref{app:sma}), sits exactly at $\omega_t=\JAB$. However, in the measured DSF this energy falls in the intensity dip between the dominant lower band and the faint upper maximum; the triplon therefore does not appear as a dominant spectral feature. The numerical MPS-SMA (black points) coincides with this one-triplon line at $r=1$ for a simple structural reason: acting on a product of on-bond singlets, a local excitation operator $S^z$ creates an on-bond triplet, so the $2L$-dimensional variational space collapses onto the $L$ single-triplet states, which do not couple across bonds at $r=1$. The SMA therefore returns the flat triplon at $\omega=\JAB$, while the TDVP spectrum contains substantial spectral weight both below and above this energy.

\paragraph{Away from the dimer point ($r\neq1$).}

For $r=0.72$ and $r=1.38$ [Fig.~\ref{fig:headline}(a),(b) and (e),(f)], the dominant low-energy bands acquire dispersion while retaining the same overall continuum structure seen at the VBS point. At $r=0.72$, the minimum of the bright low-energy band is commensurate, sitting at the zone centers $q=0$, $2\pi$, and $4\pi$. This low-energy arc can be viewed as the gapped precursor of the des~Cloizeaux--Pearson two-spinon continuum of the uniform (zigzag) Heisenberg chain~\cite{dCP_1962,Muller_1981}.

By contrast, at $r=1.38$ the low-energy response becomes incommensurate. The dominant intensity develops near $q\approx1.2\pi$ and $2.8\pi$, with much fainter mirror features near $q\approx0.83\pi$ and $3.17\pi$. Interestingly, this dynamical incommensurability is already anticipated in the ground state: the apical static structure factor peaks at the incommensurate $k\approx0.78\pi$ [Fig.~\ref{fig:ssf}(a)], while the basal channel remains locked near $\pi$. We show below that the same sublattice selectivity emerges naturally from the two-spinon description of the dynamical excitations.

The upper part of the spectrum follows the same qualitative pattern in both regimes. The faint secondary maximum remains separated from the dominant low-energy band by an intensity dip near $\omega\simeq\JAB$, reproducing the same three-feature structure found at the exact VBS point, but now with finite dispersion. As at the VBS point, the expected one-triplon dispersion (red) does not track any of the prominent features of the measured DSF; instead, the spectral weight in its vicinity remains suppressed. The MPS-SMA modes (black points) plunge below the triplon for $r\neq1$; once the ground state is dressed, the local probes acquire overlap with low-lying excitations, but the resulting branch does not provide a dispersion estimate for any of the prominent bands observed in the DSF.

The common continuum structure across all three regimes therefore calls for a description beyond local triplon excitations. We now show that it arises naturally from deconfined spin-$1/2$ domain walls of the dimerized ground state.

\subsection{Spinon origin of the dynamical continuum}
\label{sec:spinon}

\begin{figure}[t]
  \centering
  \includegraphics[width=\columnwidth]{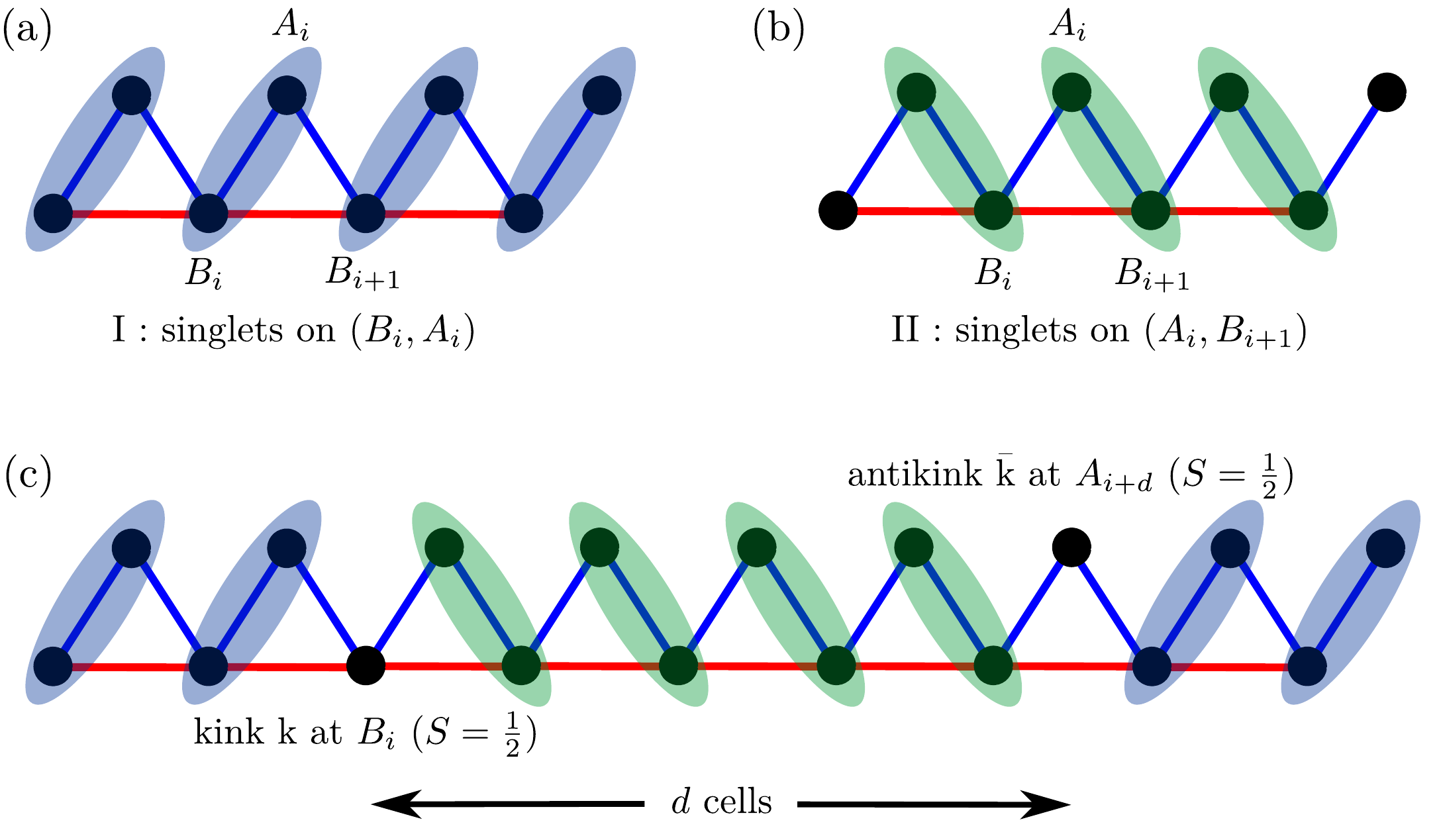}
  \caption{\label{fig:cartoon}%
    Elementary excitations over the exact VBS state at $r=1$. Away from $r=1$, the ground state retains the same bond order on average (Sec.~\ref{sec:gs}) but is no longer a tensor product of singlets. [(a)-(b)]~The two degenerate singlet coverings of the sawtooth chain at $r=1$: $\mathrm{I}$ pairs each apical spin with the basal spin of its own cell $(B_i,A_i)$; $\mathrm{II}$ with that of the next cell $(A_i,B_{i+1})$. (c)~The two-spinon pair state $\ket{i,d}$. Between the two domain walls
    the covering flips from $\mathrm{I}$ to $\mathrm{II}$. The left wall located at $i$ is a kink $\rm k$: an orphaned basal spin $B_i$ whose triangle retains its singlet. The right wall located at $i+d$ is an
    antikink $\bar{\rm k}$: an orphaned apical spin $A_{i+d}$
    whose triangle has lost its singlet. Each wall
    carries $S=\tfrac12$, and the $\Delta S=1$ of a local spin flip due to $S^z$ fractionalizes into the pair. Here $d$ is the kink--antikink separation in
    cells. Because the covering-$\mathrm{II}$ domain between the walls costs no energy, the pair potential $V(d)=0$ at every separation: the pair is
    deconfined. The coincident case $d=0$ is the on-bond triplet.}
\end{figure}

To identify the microscopic origin of the continuum, it is useful to begin at the exactly solvable VBS point, where the elementary excitations admit a particularly transparent real-space description. The two degenerate apex--base singlet coverings [Fig.~\ref{fig:cartoon}(a),(b)] support domain walls carrying spin $S=\tfrac12$, conventionally referred to as kink and antikink spinons. The kink is an orphaned basal spin whose triangle retains its singlet, while the antikink is an orphaned apical spin whose triangle has lost its singlet [Fig.~\ref{fig:cartoon}(c)]. A local $S^z$ excitation carries $\Delta S=1$ and therefore creates these domain walls in pairs. The corresponding pair state $\ket{i,d}$ contains a kink $\rm k$ at cell $i$ and an antikink $\bar{\rm k}$ at cell $i+d$. Crucially, separating the two domain walls does not produce an energy cost that grows with their separation. The intervening region simply switches from covering $\mathrm{I}$ to the exactly degenerate covering $\mathrm{II}$, leaving the number of intact singlets unchanged at $L-1$. The diagonal energy is therefore
\begin{equation}
  \bra{i,d}(H-E_0)\ket{i,d}=\tfrac34\JAB+\tfrac14\JAB\,\delta_{d,0},
  \label{eq:Vd}
\end{equation}
the $d$-independent rest energy of the one dissolved singlet plus a contact term for the on-bond triplet. The pair potential therefore vanishes at every separation, $V(d)=0$\footnote{Only singlet bonds contribute to the diagonal because, on a product state, any bond that is not itself a singlet has zero expectation:
$\expval{\mb S_i\!\cdot\!\mb S_j}=\expval{\mb S_i}\!\cdot\!\expval{\mb S_j}=0$,
since $\expval{\mb S}=0$ for a spin inside a singlet. In particular, the
base-base bonds, which never host a singlet, drop out of the diagonal
at every $r$, not only at the VBS point. They act only
off-diagonally, where they generate the wall hopping at $r\neq1$.}. This is analogous to the deconfinement mechanism for the soliton spinons of the Majumdar--Ghosh chain, whose two dimer coverings are also exactly degenerate~\cite{ShastrySutherland_1981}. It contrasts with confined domain-wall pairs (mesons) bound by a growing potential in quantum spin chains~\cite{Kormos_2016,Liu2019,Ranabhat_2025} and with two-dimensional valence-bond solids, where a separated pair generally drags a string of disrupted dimers and is confined into a triplon~\cite{ReadSachdev_1989,ReadSachdev_1990}.

The spinon picture can be tested against the TDVP spectrum at two complementary levels. First, the single-kink and single-antikink dispersions determine the kinematic support of the two-spinon continuum and can therefore be compared directly with the dispersion observed in $S^{zz}(q,\omega)$. Second, the kinematic boundaries alone do not determine how the spectral weight is distributed within the continuum. We address this separately in Sec.~\ref{sec:b1} by constructing the kink--antikink pair explicitly with a finite maximum separation.

Because the kink and antikink are deconfined, their pair energy is additive, and the continuum is given by the cross-convolution of the two spinon bands,
\begin{equation}
  \omega(q)=\varepsilon_{\rm k}(k_1)+\varepsilon_{\bar{\rm k}}(q-k_1),
  \label{eq:convol}
\end{equation}
with the internal momentum $k_1$ free (derived in Appendix~\ref{app:twospinon}). At the VBS point, $\varepsilon_{\rm k}\equiv0$ and $\varepsilon_{\bar{\rm k}}(k)=(\tfrac54+\cos k)\JAB$, so $\omega(q)=(\tfrac54+\cos(q-k_1))\JAB$ sweeps the full antikink band $[\tfrac14,\tfrac94]\JAB$ at every $q$. The flatness is therefore a direct consequence of the dispersionless kink: it absorbs the total momentum at zero energy cost and frees the antikink to sit anywhere in its band, so the antikink bandwidth survives as the full width $2\JAB$ of a momentum-independent continuum.

This closed-form continuum is precisely what is observed numerically. Its lower edge, flat at $\omega=\tfrac14\JAB$, runs through the dominant bright band of the measured DSF [thick green line in Fig.~\ref{fig:headline}(d)], identifying the dominant low-energy feature as the lower edge of the two-spinon continuum. The upper edge at $\tfrac94\JAB$ bounds the visible weight from above, while the faint secondary maximum at $\omega\simeq\tfrac74\JAB$ lies within the same continuum. Thus the dominant lower band, the faint upper maximum, and the dip between them all lie within the kinematic band $[\tfrac14,\tfrac94]\JAB$. In particular, the one-triplon energy $\omega=\JAB$ lies inside the continuum rather than defining a distinct excitation branch.

Away from the exact VBS point, the base-base bonds turn on hopping of the domain walls, and both spinons acquire dispersion. The resulting closed-form dispersions (Appendix~\ref{app:twospinon}) are
\begin{align}
  \varepsilon_{\rm k}(k)        &= E_K-(\JBB-\JAB)\,\frac{4+5\cos k}{5+4\cos k},
    \label{eq:ekmain}\\
  \varepsilon_{\bar{\rm k}}(k)  &= \varepsilon_{\rm k}(k)+\tfrac12\JBB\,(1+2\cos k),
    \label{eq:eakmain}
\end{align}
where the additive constant $E_K$ is fixed by the exact-diagonalization spectral gap (Appendix~\ref{app:twospinon}), an anchor that at $r=1$ recovers the expected value $\simeq0.21\JAB$~\cite{NakamuraKubo}. At $r=1$, Eqs.~\eqref{eq:ekmain} and \eqref{eq:eakmain} reduce to the dimer limit $\varepsilon_{\rm k}=0$ and $\varepsilon_{\bar{\rm k}}=(\tfrac54+\cos k)\JAB$.

Inserting Eqs.~\eqref{eq:ekmain} and \eqref{eq:eakmain} into the convolution \eqref{eq:convol} generates the two-spinon continua for $r=0.72$ and $r=1.38$ [Fig.~\ref{fig:headline}(b),(f)]. The central observation at the dimer point carries over throughout the dimerized phase: the lower edge of the two-spinon continuum (thick green) traces the dominant low-energy band of the measured DSF. As $r$ changes, the minima of these bands migrate across the zone, and the closed-form continuum edge reproduces this migration.

At $r=0.72$, the minimum remains commensurate, sitting at the zone centers $q=0$, $2\pi$, and $4\pi$. Although the low-energy arc is the gapped precursor of the des~Cloizeaux--Pearson two-spinon continuum of the uniform Heisenberg chain~\cite{dCP_1962,Muller_1981}, quantitatively its shape is still described by the kink--antikink edge of Eqs.~\eqref{eq:ekmain} and \eqref{eq:eakmain}, as seen in Fig.~\ref{fig:headline}(b).

By contrast, at $r=1.38$ the band minimum is incommensurate, and its position follows directly from the spinon dispersions. The kink band is stationary only at $k=0$ and $k=\pi$, so the incommensurability is carried entirely by the antikink. The curvature of Eq.~\eqref{eq:eakmain} at $k=\pi$ is $9\JAB-8\JBB$, which changes sign at $r=\tfrac98$. For $1\leq r\leq\tfrac98$, the antikink minimum is locked to the commensurate point $k=\pi$, while for $r>\tfrac98$ it moves continuously according to
\begin{equation}
\cos k^{*}=\frac{3\sqrt{1-1/r}-5}{4}.
\end{equation}
At $r=1.38$, $k^{*}=0.827\pi$, together with its reflection $2\pi-k^{*}=1.173\pi$, giving an exactly degenerate pair of minima flanking $q=\pi$, with $q=\pi$ itself a shallow local maximum between them ($\lesssim0.01\JAB$ throughout the phase). In the extended zone $q\in[0,4\pi]$, this produces four minima $q=2\pi n\pm k^{*}$ [Fig.~\ref{fig:headline}(f)]. The two minima on either side of $q=\pi$ (and their mirrors about $q=3\pi$) are not equivalent in intensity: the apex form factor $\tfrac12(1-\cos\tfrac q2)$ favors the one closer to $q=2\pi$. We therefore observe brighter intensity in the DSF near $q\approx1.2\pi$ and $2.8\pi$ [Fig.~\ref{fig:headline}(e),(f)], while the mirror partners near $q\approx0.83\pi$ and $3.17\pi$ remain faint. This also explains the sublattice selectivity noted above: the antikink, which is the apical domain wall, carries the incommensurability, while the kink, which is the basal domain wall, remains commensurate.

The same kinematics organizes the upper part of the spectrum. In both regimes, the faint secondary maximum lies inside the continuum, below the upper edge that bounds the visible weight, with the intensity dip near $\omega\simeq\JAB$ between it and the dominant band. The closed-form spinon dispersions therefore determine where the continuum lies in momentum and energy. To understand why the spectral weight is distributed within it as observed in TDVP, we now turn to an explicit real-space construction of the kink--antikink pair.

\subsection{Spectral weight and triplon fractionalization}
\label{sec:b1}

\begin{figure*}[t]
  \centering
  \includegraphics[width=\textwidth]{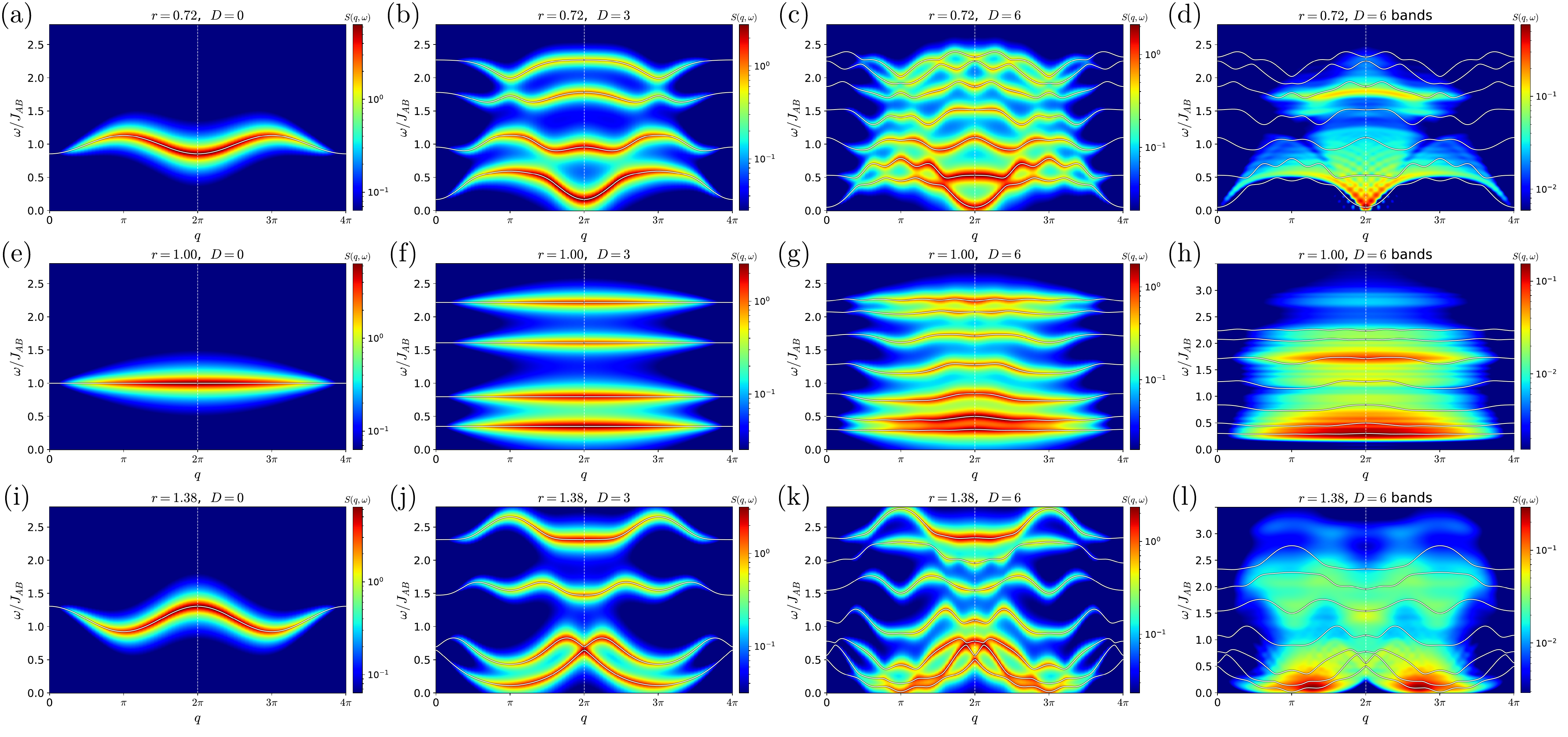}
  \caption{\label{fig:b1}%
    Finite-$D$ two-kink structure factor and its convergence to the numerical
    DSF. Rows: $r=0.72$ (a-d), $1.000$ (e-h), $1.38$ (i-l). Columns 1-3:
    the two-kink DSF at $D=0$ (coincident pair $=$ triplon), $D=3$, and $D=6$,
    showing the single triplon band splitting and delocalizing into a set of
    $D{+}1$ bands that densify from $\omega=\JAB$ downward toward the two-spinon
    gap as the pair is allowed to separate. Column 4 ($D=6$ bands): the $D=6$
    band positions (white) overlaid on the $L=32$, $\chi=600$ TDVP DSF,
    tracking the two bright low-energy bands and the suppression of weight at
    the triplon energy $\omega=\JAB$. All panels use the physical extended-zone
    convention $q\in[0,4\pi]$; energies in units of $\JAB$.}
\end{figure*}

The closed-form spinon dispersions determine the kinematic support of the continuum and account for the location of its dominant low-energy edge, but they do not determine how the spectral weight is distributed within it. To address this question, we employ the finite-$D$ two-kink construction of Sec.~\ref{sec:methods-twokink}, which follows a locally created $S=1$ excitation as its two spin-$1/2$ constituents are allowed to separate up to a maximum distance $D$.

Figure~\ref{fig:b1} shows this reconstruction. At $D=0$, the spectrum consists of the single triplon band, flat at $\omega=\JAB$ for $r=1$ and dispersing for $r\neq1$ [Fig.~\ref{fig:b1}(a),(e),(i)]. As $D$ increases, the pair delocalizes, and the local $S=1$ spin flip fractionalizes into a pair of $S=\tfrac12$ constituents. The single band splits into $D+1$ bands and progressively redistributes its spectral weight toward the lower two-spinon edge. Already at $D=6$, the discrete bands densify toward a continuum whose brightest weight lies well below $\omega=\JAB$ [Fig.~\ref{fig:b1}(c),(g),(k)].

Overlaying the $D=6$ bands on the numerical DSF [Fig.~\ref{fig:b1}(d),(h),(l)] shows agreement both in the band positions and in the relative distribution of spectral weight. The reconstruction reproduces the dominant lower band, the faint upper maximum, and the intensity dip near $\omega\simeq\JAB$ in all three regimes. In particular, the fainter upper-intensity maxima are not separate excitations; their shapes and positions coincide with upper two-kink bands. At $r=1$, the flat feature near $\omega\simeq1.7\,\JAB$ matches the $D=6$ band at $1.71\,\JAB$. At $r=1.38$, the brightest upper feature ($\omega\simeq1.45\,\JAB$ at $q=2\pi$) lies on the $D=6$ band at $1.43\,\JAB$, while at $r=0.72$ the brightest upper feature ($\omega\simeq1.80\,\JAB$ at $q=2\pi$) is traced by a $D=6$ band at $1.87\,\JAB$. Thus the closed-form convolution determines where the continuum lies [Fig.~\ref{fig:headline}, green], while the finite-$D$ two-kink construction explains how the spectral weight is distributed within it [Fig.~\ref{fig:b1}]. Together, they show that the full low-energy lineshape of the DSF is generated by the deconfined kink--antikink pair.

The finite-$D$ construction also makes explicit why the bond-operator triplon, which can be quantitatively successful in two-dimensional valence-bond solids, traces no prominent feature of the sawtooth spectrum. The triplon is the $D=0$ limit of the two-kink state [Fig.~\ref{fig:b1}(a),(e),(i)], and restricting the excitation to $D=0$ is justified only when a confining potential keeps the pair bound. In two dimensions, a separated pair generally drags a string of disrupted dimers, causing $V(d)$ to grow with separation~\cite{ReadSachdev_1989,ReadSachdev_1990,Senthil_2004}; deconfined spinons instead require a different ground state, such as a quantum spin liquid~\cite{Han_2012}. The resulting confined $S=1$ quasiparticles are observed, for example, in the pinwheel valence-bond solid of two-dimensional kagome antiferromagnets, where triplon bands form the primary excitations and are quantitatively described by dimer-based theories, with continuum scattering appearing only above a higher-energy threshold~\cite{Matan_2010,Matan_2014,YangKim_2009,CalongeMartinez_2026}.

On the sawtooth chain, by contrast, $V(d)=0$ [Eq.~\eqref{eq:Vd}], and as $D$ grows the spectral weight drains from the local triplon into the continuum [Fig.~\ref{fig:b1}], leaving an intensity dip near $\omega\simeq\JAB$ where the sharpest triplon feature would be expected in a confined system. The spinon dispersions and finite-$D$ reconstruction therefore provide complementary evidence for the same physical picture: the former determine where the two-spinon continuum lies in momentum and energy, while the latter shows how the spectral weight of a locally created triplet is transferred into that continuum as the kink and antikink are allowed to separate. Together with the TDVP results, this identifies the broad dynamical response throughout the dimerized phase as the spectral signature of spinon fractionalization.

% =====================================================================
\section{Summary and outlook}
\label{sec:conclusion}

We have computed the ground-state properties and the zero-temperature dynamical structure factor of the dimerized phase of the antiferromagnetic sawtooth chain at three representative points, $r=0.72$, $1.00$, and $1.38$. We employ a $U(1)$-symmetric DMRG\,$+$\,TDVP pipeline, benchmarked against the exactly solvable valence-bond solid at $r=1$ and exact diagonalization. The bond-resolved dimer order, bipartite entanglement entropy, and sublattice-resolved static structure factor consistently characterize a dimerized phase bounded by transitions of opposite character. Approaching the continuous lower boundary, the dimer order weakens, and the entanglement grows, whereas toward the first-order upper boundary the strong dimerization persists while the apical correlations become incommensurate, anticipating the noncollinear phase beyond.

The dynamical structure factor exhibits a broad continuum throughout the dimerized phase: a dominant lower band carries the largest share of the spectral weight, while the remaining weight spreads into a much fainter continuum with a weak secondary maximum in its upper part. Remarkably, the spectral weight is suppressed in the vicinity of the expected one-triplon band rather than concentrated around it. Our central result is the identification of this response with a deconfined two-spinon continuum built from the kink and antikink domain walls between the two degenerate singlet coverings: a locally created $\Delta S=1$ excitation fractionalizes into two $S=1/2$ constituents. This assignment is supported by two complementary calculations. The spinon dispersions determine the kinematic boundaries of the continuum and track its prominent low-energy edge, including the incommensurate minima observed at $r=1.38$, while the finite-$D$ projected two-spinon calculation reproduces the redistribution of spectral weight into the continuum, including the intensity dip near the expected triplon band and the fainter upper maximum. The dimerized phase thus connects, in both its static correlations and dynamical response, the physics of the des~Cloizeaux--Pearson two-spinon continuum of the uniform Heisenberg chain on one side and the gapless noncollinear phase on the other~\cite{Rausch_2025}, with the exactly solvable VBS point providing a particularly transparent realization of the underlying spinon physics.

Two directions follow naturally. First, the sawtooth chain provides a useful building block for understanding the recently discovered class of Ti$^{3+}$ kagome fluorides, where strong modulations of the exchange interactions generate coupled sawtooth-like motifs~\cite{Thennakoon_2025,Thennakoon_unpub}. Broad magnetic continua have also been observed in the dynamical response of these materials, including Cs$_8$LiNa$_3$Ti$_{12}$F$_{48}$~\cite{Thennakoon_unpub}. The real materials, however, are two-dimensional, and their excitation spectra necessarily involve couplings between the sawtooth motifs as well as other details absent from the present one-dimensional model. An important question for future work is therefore how the deconfined spinons identified here evolve upon coupling the chains into two dimensions, and in particular whether and how inter-chain interactions confine, or weakly confine, the kink-antikink pairs into bound excitations. Second, the domain-wall picture developed here should provide a useful starting point for studying the excitations at finite magnetic field, where flat-band physics is closely connected to localized-magnon states and the magnetization plateaus and jumps of the sawtooth family~\cite{Schulenburg2002,Derzhko_2020}. We leave these questions to future work.

% =====================================================================
\begin{acknowledgments}
This work was supported by the U.S. Department of Energy, Office of Science, Basic Energy Sciences, under Award No. DE-SC0026087. The authors acknowledge Research Computing at The University of Virginia for providing computational resources that have contributed to the results reported within this publication.
\end{acknowledgments}

% =====================================================================
%  APPENDICES (merged from the former supplemental.tex, trimmed of material
%  duplicated in the main text; all SM figures and the table retained).
% =====================================================================
\appendix

\section{Tensor-network pipeline: explicit transform and error analysis}
\label{app:tn}

The DMRG\,$+$\,TDVP pipeline is specified in
Sec.~\ref{sec:methods-dsf}. Here, we explain the space-time Fourier transformation and the truncation-error diagnostics behind the convergence
of the numerical methods. The sawtooth chain is open with $N=2L$ spins at positions $r_{\sigma,j}=j+\delta_\sigma$ ($\delta_B=0$, $\delta_A=\tfrac12$). Each correlator channel $G^{\sigma'\sigma}(j,t)$ [Eq.~\eqref{eq:greens}] is
transformed with the $E_0$ phase restored and a Gaussian time window
$W(t)=e^{-t^2/2\sigma_t^2}$, $\sigma_t=t_{\rm final}/3$:
\begin{equation}
\begin{aligned}
  S^{\sigma'\sigma}(q,\omega)
  = &-\frac{1}{\pi}\,\mathrm{Im}\sum_{j}
     e^{\,iq\,(r_{\sigma',j}-r_{\sigma,j_0})}\\
  &\times\int_0^{t_{\rm final}}\!\! dt\;
     W(t)\,e^{\,i(\omega+E_0)t}\,G^{\sigma'\sigma}(j,t),
\end{aligned}
\label{eq:tn-dsf}
\end{equation}
and the total is $S^{zz}(q,\omega)=\frac1N\sum_{\sigma'\sigma}
S^{\sigma'\sigma}(q,\omega)$. The relative positions carry the apex offset,
placing the DSF in the extended-zone convention of
Sec.~\ref{sec:methods-dsf} (asymmetric about $q=\pi$, symmetric about
$q=2\pi$; cf.\ Sec.~\ref{sec:dsf}). Production parameters are as follows: $L=32$ (source
cell $j_0=16$), $\chi_{\rm DMRG}=128$, $\chi_{\rm TDVP}=600$ ($480$ for the
convergence check), $\Delta t=0.01\,\JAB^{-1}$, and
$t_{\rm final}=8\,\JAB^{-1}$.

\paragraph{DMRG convergence and ground-state accuracy.}

\begin{figure*}[t]
  \centering
  \includegraphics[width=\textwidth]{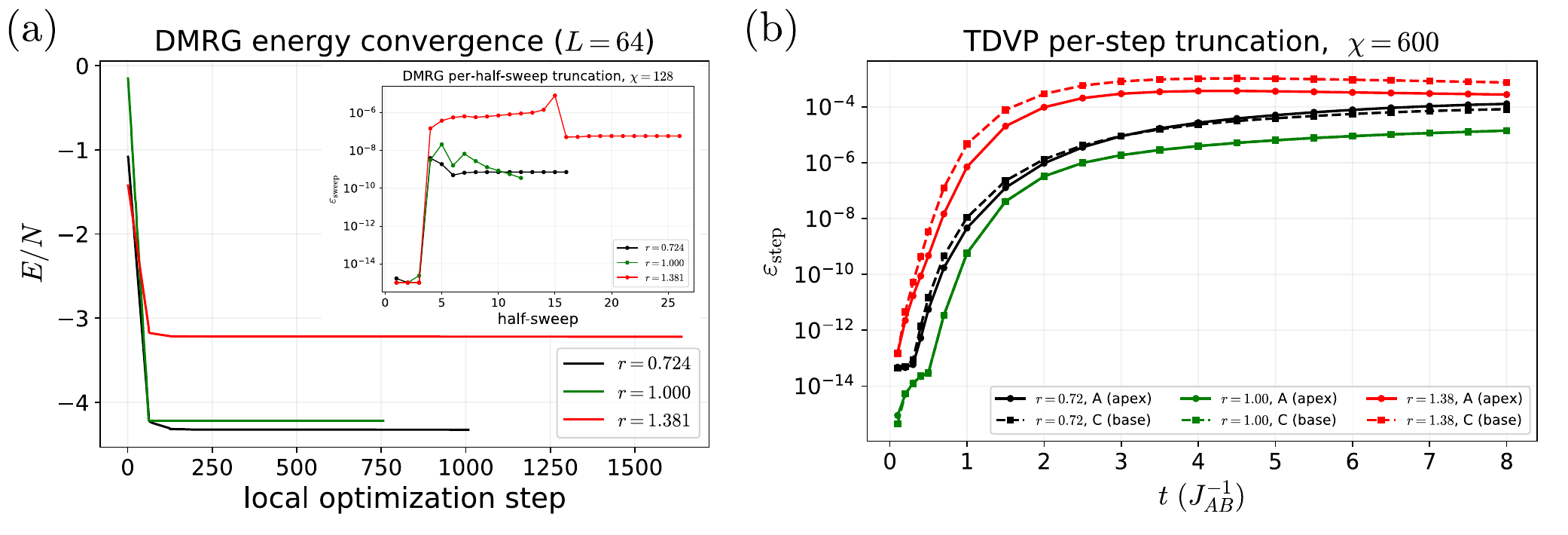}
  \caption{\label{fig:s1err}%
    Truncation-error analysis of the two tensor-network stages.
    (a)~DMRG energy convergence: energy per site $E/N$ (raw exchange units)
    versus local two-site optimization step for the three $L=64$ ground states
    at $\chi_{\rm DMRG}=128$ (63 local updates per half-sweep). The $r=1.38$
    run (red) exhibits the intermediate metastable plateau discussed in the
    text before dropping onto the true ground state. Inset: discarded Schmidt
    weight per half-sweep $\varepsilon_{\rm sweep}$; the truncation is at
    machine zero during the adaptive growth phase ($\chi<128$), spikes to
    $8\times10^{-6}$ at the half-sweep where the $r=1.38$ run escapes the
    plateau, and settles to $\lesssim10^{-9}$ ($r\le1$) and
    $\sim5\times10^{-8}$ ($r=1.38$) in the converged tail.
    (b)~TDVP per-step truncated weight $\varepsilon_{\rm step}(t)$ (summed
    discarded Schmidt weight of all two-site updates within one time step) at
    $\chi_{\rm TDVP}=600$ for the six production evolutions (three regimes
    $\times$ apex/basal source). The truncation rises during the initial
    entanglement growth, saturates once the bond dimension reaches its cap,
    and is ordered by the distance from the exact VBS point
    ($r=1$ lowest, $r=1.38$ highest).}
\end{figure*}

In Table~\ref{tab:dmrg-quality} we present the convergence record of the six production ground states ($L=32$ for seeding the dynamics, $L=64$ for the equilibrium characterization). The largest bond dimension was reached in all $r \neq 1$ cases, signifying the presence of truncation errors, the number of sweeps to convergence (sweeping continues until the energy
per site is stable to the tolerance of $10^{-7}$ over a sliding window of ten half-sweeps),
and the energy variance per site,
\begin{equation}
  v \;\equiv\; \frac{\langle H^2\rangle-\langle H\rangle^2}{N\,\JAB^2},
  \label{eq:variance}
\end{equation}
contracted exactly as a two-MPO-layer sandwich on the converged MPS ($v=0$
for an exact eigenstate). A fixed bond dimension can bias DMRG toward metastable solutions; the sweep histories provide a direct check
[Fig.~\ref{fig:s1err}(a)]. $r=1.38$ is the most difficult case of convergence among all three values of $r$ considered. At $r=1.38$, $L=64$, the energy passes through an
intermediate plateau before falling onto the true ground state, after which it is stable to $\sim3\times10^{-11}$ per site for the next ten half sweeps concluding the simulation. The other two $r$ follow a similar pattern but are converged at a lesser number of sweeps with a few orders of magnitude smaller $v$. The exactly solvable case of $r=1$ is reproduced by DMRG with a bond dimension of 4, thus implying the tensor product VBS state.

\begin{table}[t]
  \centering
  \caption{\label{tab:dmrg-quality}%
    DMRG convergence and accuracy of the six production ground states.
    $\chi_{\rm fin}$ is the largest bond dimension actually reached
    ($\chi_{\rm DMRG}=128$ allowed), ``sweeps'' the number of full sweeps to
    convergence, and $v$ the energy variance per site of
    Eq.~\eqref{eq:variance} in units of $\JAB^2$ (entries $\lesssim10^{-14}$
    are at machine precision).}
  \begin{ruledtabular}
  \begin{tabular}{llcccc}
    $r$ & $L$ & $E_0/N\JAB$ & $\chi_{\rm fin}$ & sweeps & $v$ \\
    \colrule
    $0.72$ & $32$ & $-0.384133$ & $128$ & $7$  & $2.5\times10^{-12}$ \\
    $1.00$ & $32$ & $-0.375000$ & $4$  & $6$  & $4\times10^{-15}$   \\
    $1.38$ & $32$ & $-0.394877$ & $128$ & $7$  & $4.6\times10^{-9}$  \\
    $0.72$ & $64$ & $-0.384424$ & $128$ & $8$  & $6.0\times10^{-11}$ \\
    $1.00$ & $64$ & $-0.375000$ & $4$   & $6$  & $7\times10^{-15}$   \\
    $1.38$ & $64$ & $-0.395235$ & $128$ & $13$ & $5.8\times10^{-9}$  \\
  \end{tabular}
  \end{ruledtabular}
\end{table}

\begin{figure*}[t]
  \centering
  \includegraphics[width=\textwidth]{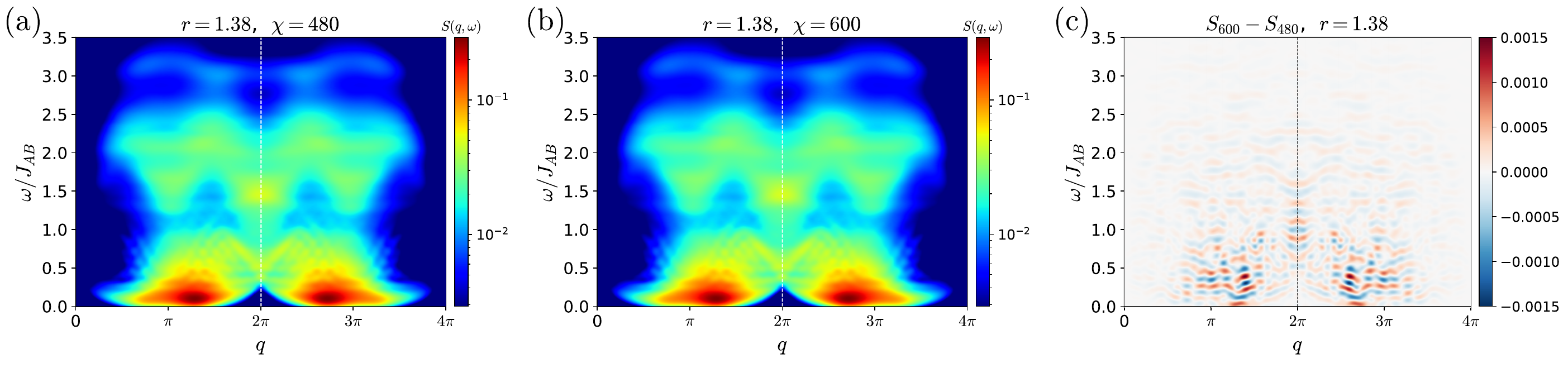}
  \caption{\label{fig:s1chi}%
    End-to-end bond-dimension convergence of the DSF in the least favorable
    regime, $r=1.38$. (a),(b)~$S^{zz}(q,\omega)$ from the full
    DMRG\,$+$\,TDVP pipeline repeated at $\chi_{\rm TDVP}=480$ (a) and $600$
    (b), on a common logarithmic color scale. (c)~Pointwise difference
    $S_{600}-S_{480}$ on a linear scale: the two maps agree to $0.6\%$ in
    relative $L^2$ norm, with a maximum pointwise deviation of $0.7\%$ of the
    peak intensity.}
\end{figure*}

\paragraph{TDVP error control.}
The real-time evolution uses the two-site TDVP integrator, so the bond
dimension grows adaptively with the entanglement of $|\phi^\sigma(t)\rangle$
up to $\chi_{\rm TDVP}=600$. The error is monitored through the per-step truncated weight $\varepsilon_{\rm step}(t)$, the summed
discarded Schmidt weight of all two-site updates within a single time step [Fig.~\ref{fig:s1err}(b)]. It rises steeply during the initial entanglement growth of the propagating excitation, saturates once the bond dimension reaches its cap. At
$t_{\rm final}=8\,\JAB^{-1}$, $\varepsilon_{\rm step}\simeq O(10^{-5})$ at
$r=1$, and $\simeq O(10^{-4})$ at $r\neq 1$ (the two values per regime are the two source channels). As an end-to-end convergence test in the least favorable regime, the entire pipeline was repeated at $r=1.38$ with $\chi_{\rm TDVP}=480$ to compute the relative error in the DSF. The resulting $S^{zz}(q,\omega)$ differs from the
$\chi_{\rm TDVP}=600$ result by $0.6\%$ in relative $L^2$ norm, with a
maximum pointwise deviation of $0.7\%$ of the peak intensity, i.e.\ the DSF maps shown in the main text are converged in $\chi$ on the scale of the plots (Fig.~\ref{fig:s1chi}).

% ---------------------------------------------------------------------
\section{Single-mode approximation}
\label{app:sma}

\paragraph{Variational framework.}
For a set of local excitation operators $\{\Omega^\sigma_j\}$ acting on a
ground state $|\Psi_0\rangle$ (energy $E_0$), the trial states
$\Omega^\sigma_j|\Psi_0\rangle$ are generally non-orthogonal, so the
variational spectrum follows from the generalized eigen-problem
[Eq.~\eqref{eq:sma-main} of the main text]
\begin{align}
  \tilde H\,\psi &= \omega\,\mathcal O\,\psi, \label{eq:sma-app}\\
  \mathcal O^{\sigma\sigma'}_{ij} &=
    \langle\Psi_0|(\Omega^\sigma_i)^\dagger\Omega^{\sigma'}_j|\Psi_0\rangle, \\
  \tilde H^{\sigma\sigma'}_{ij} &=
    \langle\Psi_0|(\Omega^\sigma_i)^\dagger(H-E_0)\Omega^{\sigma'}_j|\Psi_0\rangle .
\end{align}
Equation~\eqref{eq:sma-app} is the stationarity condition of the
Rayleigh--Ritz principle restricted to the trial subspace: extremizing
$R(\psi)=\psi^\dagger\tilde H\psi/\psi^\dagger\mathcal O\psi$ over the
coefficients yields Eq.~\eqref{eq:sma-app} with $\omega=R(\psi)$, the metric
$\mathcal O$ being the non-orthogonality of the trial states. The trial subspace is a subspace of the full Hilbert space, each
generalized eigenvalue is a variational upper bound on the true excitation of the same rank. For a single operator this collapses to the single-mode Rayleigh quotient ~\cite{Sharma_2025}. The multi-operator (matrix) form used here lets a small set of local probes mix variationally.

\paragraph{One-bond triplon branch.}
The $S^z$, $\Delta S^z_{\rm tot}=0$ singlet-to-triplet operator on the
apex--base bond, $\Omega^{\rm mag}_j=S^z_{B_j}-S^z_{A_j}$, gives to one-bond
order
\begin{equation}
  \omega_{\rm mag}(k)=\JAB+\tfrac{\JBB-\JAB}{2}\cos k ,
  \label{eq:magnon}
\end{equation}
flat at $\omega=\JAB$ at the VBS point $r=1$, with band minimum migrating
from $k=0$ ($r<1$) to $k=\pi$ ($r>1$). This is the SMA realization of the
single-triplet mode and coincides with the self-consistent bond-operator
triplon of Appendix~\ref{app:bondop}; the two analytic routes agree across the dimerized phase. Within the numerical basis $\{S^z_{B_j},S^z_{A_j}\}$ used below, this triplon is the antisymmetric combination $S^z_{B_j}-S^z_{A_j}$; the symmetric combination $S^z_{B_j}+S^z_{A_j}$ builds the total $S^z$ and annihilates the singlet ground state, a spurious
null mode removed below.

\paragraph{Numerical (MPS) SMA.}
Beyond the analytic one-bond estimate, we evaluate Eq.~\eqref{eq:sma-app}
numerically on the DMRG ground state, with the trial space spanned by the two
one-site probes per cell, $\{\Omega^{B}_j=S^z_{B_j},\,
\Omega^{A}_j=S^z_{A_j}\}_{j=1}^{L}$ ($2L$ states). The procedure is as follows:
\begin{enumerate}[leftmargin=*]
\item \textbf{Matrix elements.} Explicitly build the $2L\times2L$ matrices $\mathcal O^{\sigma\sigma'}_{ij}$ and $\tilde H^{\sigma\sigma'}_{ij}$ as MPS-MPO-MPS sandwhiches, $\langle\Psi_0|(\Omega^\sigma_i)^\dagger\,\hat O\,\Omega^{\sigma'}_j|\Psi_0\rangle$
with $\hat O\in\{\mathbbm{1},H\}$.
\item \textbf{Generalized solve.} After Hermitian symmetrization, the
near-singular metric is regularized by a small diagonal shift
$\mathcal O\to\mathcal O+\varepsilon\,\mathbbm{1}$ ($\varepsilon=10^{-10}$)
and the positive-definite generalized symmetric eigen-problem is solved with
LAPACK \texttt{hegvd} (internally a Cholesky of
$\mathcal O+\varepsilon\mathbbm 1$).
\item \textbf{Momentum projection.} Each eigenvector is projected onto the
cell-momentum plane waves
$|k,\sigma\rangle=L^{-1/2}\sum_j e^{-ikj}\,\Omega^\sigma_j|\Psi_0\rangle$ on
the grid $k=2\pi m/L$; its effective momentum is the $k$ of maximal weight.
\item \textbf{Null-mode filtering.} On the singlet ground state
$\sum_j(S^z_{B_j}+S^z_{A_j})|\Psi_0\rangle=S^z_{\rm tot}|\Psi_0\rangle=0$,
producing a spurious $\omega\!\approx\!0$ eigenvalue---a Goldstone of the
$U(1)$ total-$S^z$ conservation, not a physical excitation; eigenvalues with
$|\omega|<10^{-3}$ are discarded.
\end{enumerate}
The lowest non-null eigenvalue at each $k$ is the variational SMA
dispersion. Validation: at the exact VBS ($r=1$) the method returns
the flat triplon at $\omega=\JAB$, reproducing Eq.~\eqref{eq:magnon}.

\paragraph{Interpretation of the lowest branch.}
At the exact VBS point the ground state is a tensor product of on-bond singlets, on which a local $S^z$ creates only the on-bond triplet. The kink and antikink are non-local domain walls, invisible to any one-site operator, so the MPS-SMA necessarily returns the flat triplon at $\omega=\JAB$, coinciding with the one-bond result~\eqref{eq:magnon}. Away from the VBS point, the dressed ground state lets the same local probes overlap with the much lower kink-antikink pair continuum, and the lowest branch plunges below the triplon. The lowest MPS-SMA branch is therefore a loose variational upper bound on the two-spinon continuum, not a triplon estimate. The agreement with the triplon at $r=1$ reflects the operator redundancy of the exact VBS, not a triplon character of the low-energy spectrum.

% ---------------------------------------------------------------------
\section{Bond-operator triplon theory}
\label{app:bondop}

We briefly summarize the harmonic bond-operator
calculation~\cite{SachdevBhatt} used as a local-triplon reference. We choose
the apex--base dimer covering that is exact at the VBS point
$r = \JBB/\JAB = 1$. On each unit cell $j$, we introduce a singlet boson
$s_j^\dag$ and triplet bosons $t_{j\alpha}^\dag$, where $\alpha = x,y,z$,
subject to
\begin{equation}
    s_j^\dag s_j + \sum_\alpha t_{j\alpha}^\dag t_{j\alpha} = 1.
\end{equation}
For the dimer $(A_j,B_j)$, the spin operators are written as
\begin{equation}
    \mb{S}_j^A = \frac{1}{2}(\mb{L}_j + \mb{Q}_j),\quad \mb{S}_j^B = \frac{1}{2}(-\mb{L}_j + \mb{Q}_j),
\end{equation}
with
\begin{equation}
    L_j^\alpha = s_j^\dag t_{j\alpha} + t_{j\alpha}^\dag s_j,\quad Q_j^\alpha = -i\sum_{\beta\gamma}\epsilon_{\alpha\beta\gamma}t_{j\beta}^\dag t_{j\gamma}.
\end{equation}
Condensing the singlet $s_j,s_j^\dag\to\bar s$ and enforcing the constraint
on average with a Lagrange multiplier $\lambda$, the quadratic Hamiltonian is
\begin{equation}
    H_2 = \mu\sum_{j,\alpha}t_{j\alpha}^\dag t_{j\alpha} + \frac{(\JBB - \JAB)\bar s^2}{4}\sum_{j,\alpha}T_{j\alpha}T_{j+1,\alpha},
\end{equation}
where
\begin{equation}
    \mu = \frac{\JAB}{4} + \lambda,\quad T_{j\alpha} = t_{j\alpha}^\dag + t_{j\alpha}.
\end{equation}
After Fourier transforming with
$t_{j\alpha} = L^{-1/2}\sum_ke^{ikj}t_{k\alpha}$, we obtain
\begin{equation}
    H_2 = \sum_{k,\alpha}A_kt_{k\alpha}^\dag t_{k\alpha} + \frac{1}{2}\sum_{k,\alpha}B_k\qty(t_{k\alpha}^\dag t_{-k,\alpha}^\dag + t_{k\alpha}t_{-k,\alpha}),
\end{equation}
where
\begin{align}
    A_k &= \mu + \frac{(\JBB - \JAB)\bar s^2}{2}\cos k, \\
    B_k &= \frac{(\JBB - \JAB)\bar s^2}{2}\cos k.
\end{align}
The Bogoliubov transformation
$t_{k\alpha} = u_kb_{k\alpha} + v_kb_{-k,\alpha}^\dag$, $u_k^2 - v_k^2 = 1$,
gives the harmonic triplon dispersion
\begin{equation}
    \omega_k = \sqrt{A_k^2 - B_k^2} = \sqrt{\mu[\mu + (\JBB - \JAB)\bar s^2\cos k]}.
\end{equation}
The parameters $\bar s$ and $\lambda$ are fixed self-consistently by
minimizing the mean-field ground-state energy
\begin{equation}
    E_{\rm MF} = L\qty[-\frac{3\JAB}{4}\bar s^2 + \lambda(\bar s^2 - 1)] + \frac{3}{2}\sum_k(\omega_k - A_k),
\end{equation}
which gives
\begin{gather}
    \bar s^2 + \frac{3}{2L}\sum_k\qty(\frac{A_k}{\omega_k} - 1) = 1, \\
    \lambda - \frac{3\JAB}{4} + \frac{3(\JBB - \JAB)}{4L}\sum_k\qty(\frac{\mu}{\omega_k} - 1)\cos k = 0.
\end{gather}
At the VBS point $\JBB = \JAB = J$, these equations give $\bar s^2 = 1$,
$\lambda = 3J/4$, and $\mu = J$, so the harmonic triplon is flat at
$\omega_k = J$.

The one-triplon contribution to the dynamical structure factor follows from
the linear part of the spin operator. With $r_{B,j} = j$ and
$r_{A,j} = j + \tfrac12$, the longitudinal spin operator is
\begin{equation}
    S_k^z = \frac{1}{\sqrt{2L}}\sum_je^{-ikj}\qty(e^{-ik\delta_B}S_j^{B,z} + e^{-ik\delta_A}S_j^{A,z}).
\end{equation}
Keeping only terms linear in $t_{j\alpha}$,
\begin{equation}
    S_k^z = \frac{\bar s}{\sqrt{2}}F(k)\qty(t_{kz} + t_{-k,z}^\dag),
\end{equation}
where the dimer form factor is
\begin{equation}
    F(k) = \frac{1}{2}\qty(e^{-ik\delta_A} - e^{-ik\delta_B}).
\end{equation}
Using $(u_k + v_k)^2 = (A_k - B_k)/\omega_k = \mu/\omega_k$, the
longitudinal one-triplon contribution is
\begin{equation}
    S_{\rm 1t}^{zz}(k,\omega) = \frac{\bar s^2}{2}|F(k)|^2\frac{\mu}{\omega_k}\delta(\omega-\omega_k).
\end{equation}

% ---------------------------------------------------------------------
\section{Kink--antikink (two-spinon) theory of the low-energy continuum}
\label{app:twospinon}

The prominent low-energy feature of the DSF is not the triplon of
Appendix~\ref{app:bondop} but a two-spinon continuum built from the two degenerate valence-bond-solid (VBS) coverings. We give the construction in full, following the dressed-domain-wall approach~\cite{HaoTchernyshyov}. It is the deconfined counterpart of the two-kink meson problem~\cite{Liu2019}: a constrained two-wall Hilbert space carrying a non-orthogonal metric, an effective hopping Hamiltonian.

\subsection{Domain walls and the non-orthogonal metric}
At $r=1$ the exact ground state is a product of apex--base singlets on one of
two coverings $\ket{\mathrm{I}},\ket{\mathrm{II}}$. Its elementary
excitations are the two domain walls between them: a kink (an orphaned
basal spin, whose triangle still carries a singlet, so it costs no energy and
is dispersionless) and an antikink (an orphaned apical spin, whose
triangle loses its singlet, costing $\tfrac34\JAB$). A single wall is a
product state that differs from a partner displaced by $d$ cells only on the
$2d{+}1$ sites separating the two orphan positions; elsewhere the two
coverings coincide and overlap unity. On this ``domino'' the two coverings
pair the intervening spins into staggered singlets, and the free end fixes a unique
alternating spin string, each mismatched singlet contributing a factor
$-\tfrac12$. The overlap is therefore a geometric cascade,
\begin{equation}
\begin{aligned}
  o(d)&\equiv\braket{\text{wall}}{\text{wall}+d}=\qty(-\tfrac12)^{|d|},\\
  o(k)&=\sum_d o(d)\,e^{ikd}=\frac{3}{5+4\cos k},
\end{aligned}
\label{eq:overlap}
\end{equation}
computable in $O(d)$ time (a transfer matrix along the domino, not a sum over
$2^d$ spin configurations). The domain-wall basis is thus non-orthogonal and carries the metric $o(k)$ (peaked at $k=\pi$,
$o=3$; minimal at $k=0$, $o=\tfrac13$). For a two-wall state, the overlap
factorizes into the two independent domino chains, up to a local correction
where the walls share a triangle.

\subsection{Effective two-wall Hamiltonian}
The Hamiltonian in Eq.~\eqref{eq:H} is a sum of nearest-neighbor bonds, and any two two-wall states share an identical singlet background away from the walls; every matrix element $\bra{i,d}(H,\mathbbm 1)\ket{i', d'}$ is fixed by a bounded neighborhood of the walls which is independent of system size. 

\paragraph{Diagonal ($V(d)=0$).}
Each intact apex-base singlet contributes $-\tfrac34\JAB$ ($L{-}1$ of them, independent of $d$). Every base-base $\JBB$ bond connects different singlets, so $\expval{\mathbf S\!\cdot\!\mathbf S}=\expval{\mathbf S}\!\cdot\! \expval{\mathbf S}=0$. The base-base coupling is inert on the diagonal at every $r$, and at $d=0$ the collapsed triangle carries a triplet contributing $\tfrac14\JAB$. Relative to the VBS energy $E_0=-\tfrac34 L\JAB$, this reproduces Eq.~\eqref{eq:Vd} of the main text, the pair potential vanishes at every separation, $V(d)=0$ for
$d\ge1$.

\paragraph{Hopping.}
The off-diagonal elements come from the exchange
$\tfrac12(S^+_iS^-_j+\mathrm{h.c.})$ acting on a bond that touches an orphan, which displaces one wall by a single cell. Each amplitude is a finite-cluster matrix element (for example, the intra-domino base--base bond that hops the kink by one cell has bare matrix element $-\tfrac38$, evaluated below).

\paragraph{Assembly.}
With a flat diagonal [Eq.~\eqref{eq:Vd}] and finite hoppings, an eigenstate
$\ket{\psi}=\sum_{i,d}\psi(i,d)\ket{i,d}$ obeys the real-space
generalized eigen-problem $(H-E_0)\psi=\omega\,\mathcal O\psi$, the
non-orthogonal metric $\mathcal O$ [Eq.~\eqref{eq:overlap}] sitting on the
right-hand side. This is the sawtooth analogue of the effective two-kink
Hamiltonian of Ref.~\cite{Liu2019}, with $V(d)=0$ in place of their confining
potential and $\mathcal O\neq\mathbbm 1$ from the domino overlaps.

\subsection{Single-spinon dispersions in closed form}
\label{app:closedform}

For arbitrary $r$, split the Hamiltonian as $H=H_{r=1}+V$, where
$H_{r=1}-E_0=\tfrac{3\JAB}{2}\sum_l P^{(l)}_{3/2}$ is the projector sum of the dimer point and
$V=(\JBB-\JAB)\sum_l\mathbf S_{b_l}\!\cdot\!\mathbf S_{b_{l+1}}$ collects the base-base deviation. The wall states are the same product states at every $r$ (they are trial states), so the dispersions of both walls, Eqs.~\eqref{eq:ekmain}-\eqref{eq:eakmain}, follow for all $r$ at once from a handful of finite-cluster matrix elements and one lattice sum. The dimer point is recovered at the end as the $V\to0$ special case, and the additive constant $E_K$ is fixed by the exact-diagonalization (ED) spectral gap.

\paragraph{Kink.}
Every triangle of a kink state stays in the spin-$\tfrac12$ sector, so the projector sum annihilates it at every $r$ and the kink matrix element is carried by $V$ alone,
\begin{equation}
\begin{split}
  \tilde h_{\rm k}(d)&\equiv\bra{{\rm k}_i}(H-E_0)\ket{{\rm k}_{i+d}}\\
  &=(\JBB-\JAB)\sum_l\bra{{\rm k}_i}\mathbf S_{b_l}\!\cdot\!
  \mathbf S_{b_{l+1}}\ket{{\rm k}_{i+d}}.
\end{split}
\label{eq:hkinkdef}
\end{equation}
At $d=0$ this vanishes, since the base-base bonds are diagonally inert [Eq.~\eqref{eq:Vd}]. The kink's diagonal energy is $\JBB$-independent at every $r$, and its dispersion lives entirely off the diagonal. At $d=1$ the
bra and ket differ only on the three-site base--apex--base domino
$(b_i,a_i,b_{i+1})$. Any bond with one end outside the domino factorizes
through a spectator singlet common to bra and ket,
$\expval{\mathbf S}_{\rm singlet}=0$, so only the single internal base--base
bond $(b_i,b_{i+1})$ survives. Contracting
$\mathbf S_{b_i}\!\cdot\!\mathbf S_{b_{i+1}}
=S^z_{b_i}S^z_{b_{i+1}}+\tfrac12(S^+_{b_i}S^-_{b_{i+1}}+\mathrm{h.c.})$
between $\ket{s_{b_ia_i}}\!\ket{\uparrow_{b_{i+1}}}$ and
$\bra{\uparrow_{b_i}}\!\bra{s_{a_ib_{i+1}}}$ gives
$-\tfrac14-\tfrac18=-\tfrac38$, so
$\tilde h_{\rm k}(1)=-\tfrac38\,(\JBB-\JAB)$. For general separation the
domino holds $|d|$ internal base--base bonds, the edge bonds still vanish,
and by translation invariance within the domino each internal bond
contributes the same increment relative to the overlap $o(d)$; together with
the extensive singlet-sea energy, the matrix element therefore has the exact
structure
\begin{equation}
  h(d)=E_{\rm bulk}\,o(d)+\varepsilon_{\rm dom}\,|d|\,o(d)+h_{\rm conn}(d),
  \label{eq:hdstructure}
\end{equation}
with $E_{\rm bulk}\propto L$ the singlet-sea energy, $\varepsilon_{\rm dom}$ the per-cell domain increment, and $h_{\rm conn}$ a short-range connected part ($h_{\rm conn}=0$ for the kink).
The bulk-subtracted ratio
\begin{equation}
  R(d)\equiv\frac{h(d)}{o(d)}-\frac{h(0)}{o(0)}
  \label{eq:Rd}
\end{equation}
removes $E_{\rm bulk}$ exactly and is exactly linear in $|d|$; the $d=1$
element fixes the slope,
\begin{equation}
  R_{\rm k}(d)=\varepsilon_{\rm dom}\,|d|,\qquad
  \varepsilon_{\rm dom}=\frac{\tilde h_{\rm k}(1)}{o(1)}=\tfrac34\,(\JBB-\JAB).
  \label{eq:epsdom}
\end{equation}
(Diagonalizing finite windows directly instead mixes
$E_{\rm bulk}$ into the extracted bandwidth, which then drifts with window size. Eq.~\eqref{eq:Rd} is the window-free object. The linear form and the coefficients are confirmed in exact rational arithmetic against finite-cluster matrix elements on windows of $10$ and $12$ cells.)

\paragraph{Antikink.}
The antikink differs in two ways. First, its wall triangle carries
spin-$\tfrac32$ weight $\tfrac12$, so the projector survives on the
diagonal and contributes $\tfrac34\JAB\,\delta_{d,0}$, the dissolved
singlet of Eq.~\eqref{eq:Vd}, while $V$ remains diagonally inert. Second, its domino is apex-base-apex, $(a_i,b_{i+1},a_{i+1})$, with no internal base-base bond. The $V$ part vanishes at $d=1$, and a domain of width $d$ holds only $d-1$ internal bonds, so the ramp is
$\varepsilon_{\rm dom}(d-1)$. Hence
\begin{equation}
\begin{split}
  R_{\overline{\rm k}}(d)&=\varepsilon_{\rm dom}\,(d-1)-\tfrac34\JAB\\
  &=\varepsilon_{\rm dom}\,|d|+C,\qquad C=-\tfrac34\JBB,
\end{split}
\label{eq:Rdantikink}
\end{equation}
the constant combining the dissolved singlet ($-\tfrac34\JAB$) and the one missing internal bond ($-\varepsilon_{\rm dom}$), both walls disperse with the same slope.

\paragraph{Fourier transform.}
By Eq.~\eqref{eq:Rd} the matrix element factors as
$h(d)=o(d)\,[E_{\rm ref}+R(d)]$ with $E_{\rm ref}=h(0)/o(0)$ a
$d$-independent constant, so the single-wall Rayleigh quotient reads
\begin{equation}
  \varepsilon(k)=\frac{h(k)}{o(k)}
  =E_{\rm ref}+\frac{1}{o(k)}\sum_d R(d)\,o(d)\,e^{ikd}.
  \label{eq:rayleighk}
\end{equation}
The linear ramp reduces everything to a single lattice sum, evaluated with
the generating function
$o_z(k)=\sum_d z^{|d|}e^{ikd}=(1-z^2)/(1-2z\cos k+z^2)$ [which reproduces
Eq.~\eqref{eq:overlap} at $z=-\tfrac12$] and
$z\,\partial_z z^{|d|}=|d|\,z^{|d|}$:
\begin{equation}
\begin{split}
  S(k)&\equiv\sum_d |d|\,\qty(-\tfrac12)^{|d|}e^{ikd}
  =z\,\partial_z o_z(k)\Big|_{z=-\frac12}\\
  &=-\frac{4\,(4+5\cos k)}{(5+4\cos k)^2},
\end{split}
\label{eq:latticesum}
\end{equation}
so that $S(k)/o(k)=-\tfrac43\,(4+5\cos k)/(5+4\cos k)$. For the kink,
$\sum_d R_{\rm k}(d)\,o(d)\,e^{ikd}=\varepsilon_{\rm dom}S(k)$, and the prefactors combine as $\tfrac34\times\tfrac43=1$. The quotient is exactly Eq.~\eqref{eq:ekmain}, with the additive constant $E_K$ standing in for the bare reference $E_{\rm ref}$ (anchored below). For the antikink, the extra constant in Eq.~\eqref{eq:Rdantikink} adds $C\sum_{d\neq0}o(d)\,e^{ikd}=C\,[o(k)-1]$ to the sum [$R(0)=0$ by construction], i.e.\
$C\,[1-1/o(k)]=(-\tfrac34\JBB)\times[-\tfrac13(2+4\cos k)]
=\tfrac12\JBB\,(1+2\cos k)$ to the quotient, exactly the offset of
Eq.~\eqref{eq:eakmain}. Since $(4+5\cos k)/(5+4\cos k)$ runs monotonically from $+1$ at $k=0$ to $-1$ at $k=\pi$, the kink bandwidth is $2|\JBB-\JAB|=2|r-1|\JAB$, with the band minimum at $k=\pi$ for $r<1$ and at $k=0$ for $r>1$.

\paragraph{The dimer point as a special case.}
At $r=1$ the deviation $V$ switches off, so $\varepsilon_{\rm dom}=0$. Eq.~\eqref{eq:ekmain} collapses to a flat kink band and
Eq.~\eqref{eq:eakmain} to $E_K+\tfrac12\JAB\,(1+2\cos k)$. This is also the one point where the absolute reference is exact, because the bare VBS is the true ground state, and the projector Rayleigh quotient supplies it directly. Only the antikink's wall triangle carries spin-$\tfrac32$ weight [$\tilde h_{\overline{\rm k}}(0)=\tfrac34\JAB$, Eq.~\eqref{eq:Rdantikink}], so
\begin{equation}
  \omega_{\rm k}(k)=0,\qquad
  \omega_{\overline{\rm k}}(k)=\frac{\tfrac34\JAB}{o(k)}
  =\qty(\tfrac54+\cos k)\JAB,
  \label{eq:antikink-r1}
\end{equation}
this is the Nakamura-Kubo dispersion~\cite{NakamuraKubo,SSWC} with
minimum $\tfrac14\JAB$ at $k=\pi$.

\begin{figure}[t]
  \centering
  \includegraphics[width=\columnwidth]{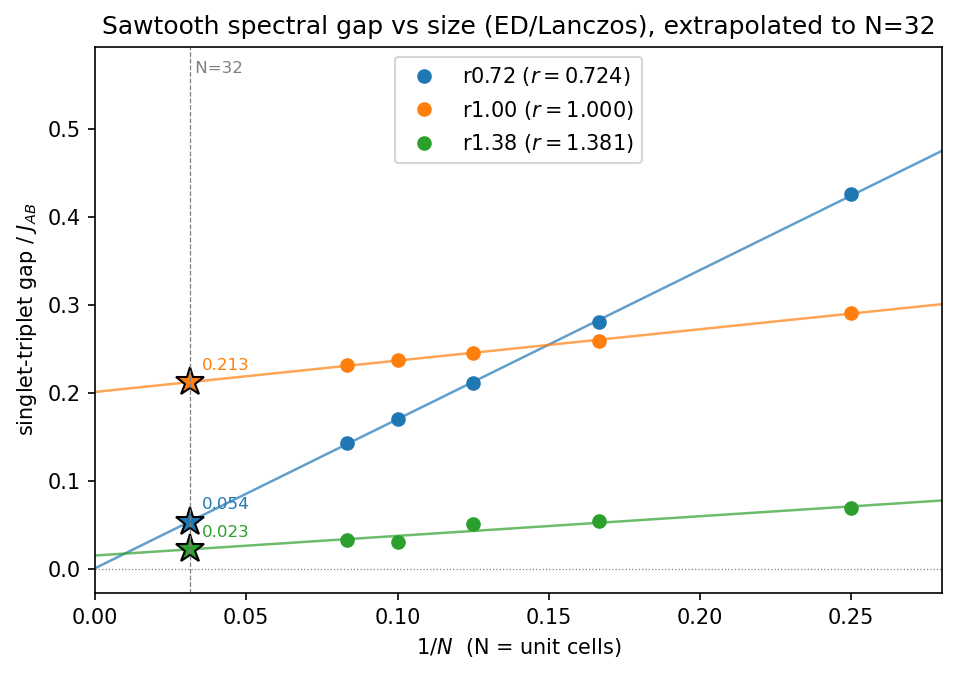}
  \caption{\label{fig:edgap}%
    ED anchor of the two-spinon continuum, Eq.~\eqref{eq:anchor}: the
    singlet--triplet spectral gap
    $E_0(S^z_{\rm tot}{=}1)-E_0(S^z_{\rm tot}{=}0)$ of the periodic sawtooth
    chain from Lanczos diagonalization at $N=4$--$12$ unit cells (circles),
    plotted against $1/N$ with linear fits $\Delta(N)=a+b/N$ (lines), for the
    three regimes. Stars mark the fits evaluated at the production size
    $N=32$ (dashed line): $\Delta_{\rm ED}=0.054$, $0.213$, and
    $0.023\,\JAB$ for $r=0.72$, $1.00$, and $1.38$, the values used to fix
    $E_K$ in Eqs.~\eqref{eq:ekmain}--\eqref{eq:eakmain}. The gap ordering
    $\Delta(1.00)>\Delta(0.72)>\Delta(1.38)$ reflects the proximity of the
    two dressed regimes to the lower and upper phase boundaries.}
\end{figure}

\paragraph{Anchoring the additive constant $E_K$ to the ED spectral gap.}
The closed forms \eqref{eq:ekmain}-\eqref{eq:eakmain} determine only energy differences, they are built entirely from the bulk-subtracted combination $R(d)$ [Eq.~\eqref{eq:Rd}], from which the entire $k$-dependence follows, while the absolute on-site reference $E_{\rm ref}=h(0)/o(0)$ does not survive the construction. For $r\neq1$ the true ground state is dressed and lies well below the bare VBS on which the wall states are built, so a wall creation energy measured
from the VBS reference is unreliable near the phase boundaries. The
excitation energies of the DSF, however, are measured from the true ground state. Since a single spinon is topological and never created alone, only the energy of the pair is observable, and the mismatch of
references is absorbed into the single global constant $E_K$, fixed by matching the bottom of the two-spinon continuum to the true spectral gap,
\begin{equation}
  \min_{q}\,\min_{k_1}\bigl[\varepsilon_{\rm k}(k_1)
  +\varepsilon_{\overline{\rm k}}(q-k_1)\bigr]=\Delta_{\rm ED}.
  \label{eq:anchor}
\end{equation}
Here $\Delta_{\rm ED}$ is the singlet--triplet gap
$E_0(S^z_{\rm tot}{=}1)-E_0(S^z_{\rm tot}{=}0)$ of the periodic chain. The lowest excitation reachable by an $S^z$ probe, i.e.\ the same object whose onset the DSF measures. The first excitation within the $S^z_{\rm tot}=0$ sector is the near-degenerate second VBS singlet, which carries no DSF weight . The gap is computed by sparse Lanczos diagonalization for rings of $N=4$-$12$ unit cells and extrapolated linearly in $1/N$ to the production size $N=32$ (Fig.~\ref{fig:edgap}). At the exactly solvable point, the anchor reproduces the known antikink gap $0.215\,\JAB$ from the ED extrapolation ~\cite{NakamuraKubo} and $0.2192\,\JAB$ from the five-cluster variational bound ~\cite{SSWC}. Everything else in the overlays (the shape of the edges, the continuum width, and the soft-mode migration) is then parameter-free.

\subsection{Two-spinon continuum}
The operator $S^z_q$ carries $\Delta S=1$ and creates the walls only in pairs. A single spinon (spin $\tfrac12$) is inaccessible within the excitation protocol. These two walls do not interact because of deconfinement, and the generalized eigen-problem $(H-E_0)\psi=\omega\,\mathcal O\psi$ factorizes. Labeling the pair by the two absolute wall positions $\ket{a,b}$ (kink at
$a$, antikink at $b$), the separated-wall overlap and Hamiltonian are both products of independent single-wall sequences,
\begin{align}
  \braket{a,b}{a',b'}&=o(a-a')\,o(b-b'),\\
  \bra{a,b}(H-E_0)\ket{a',b'}&=h_{\rm k}(a-a')\,o(b-b')\nonumber\\
  &\quad+o(a-a')\,h_{\overline{\rm k}}(b-b'),
\end{align}
with $h_{\rm k},h_{\overline{\rm k}}$ the single-wall hopping sequences. In the double-momentum basis $\ket{k_1,k_2}=\sum_{a,b}e^{ik_1 a+ik_2 b}\ket{a,b}$
both matrices are diagonal, and the fixed-$q$ generalized eigen-problem
collapses to the scalar relation
$h_{\rm k}(k_1)\,o(k_2)+o(k_1)\,h_{\overline{\rm k}}(k_2)=
\omega\,o(k_1)\,o(k_2)$. Dividing by $o(k_1)o(k_2)$, the
non-orthogonal metric cancels as a common factor leaving
$\omega=\varepsilon_{\rm k}(k_1)+\varepsilon_{\overline{\rm k}}(k_2)$ with $\varepsilon_{\rm wall}(k)=h_{\rm wall}(k)/o(k)$ the single-wall Rayleigh quotients of Eqs.~\eqref{eq:ekmain}-\eqref{eq:eakmain}. Imposing the total
momentum $q=k_1+k_2$ gives the cross-convolution of Eq.~\eqref{eq:convol}, the untruncated set of bands, with free relative motion $\psi_q(d)\sim e^{ik_1 d}$. The $q$-dependence of the continuum then reads off the closed forms, at the dimer point the flat kink [$\varepsilon_{\rm dom}=0$, Eq.~\eqref{eq:antikink-r1}] absorbs the total momentum at zero cost and the continuum sweeps the full antikink band $[\tfrac14,\tfrac94]\JAB$ at every $q$. For $r\neq1$ the dispersing kink constrains the pair and the soft mode of the lower edge migrates across the zone.

\subsection{The two-kink structure factor}
To obtain the distribution of spectral weight inside the continuum, we do a bruteforce diagonalization of the projected problem in the truncated two-wall space $\{\ket{i,d}\}$ ($i$ = kink cell, $d$ = kink-antikink separation, $0\le d\le D$). At fixed $q$ this is the generalized eigen-problem
\begin{equation}
  H(q)\,\psi_n=\omega_n\,\mathcal O(q)\,\psi_n,
  \label{eq:geneig}
\end{equation}
and, with the $d=0$ magnon as the source, the structure factor is
\begin{equation}
  S^{zz}(q,\omega)=\sum_n\big|\bra{d{=}0}\mathcal O(q)\ket{\psi_n}\big|^2\,
                   \delta\!\qty(\omega-\omega_n).
  \label{eq:2kink-sf}
\end{equation}
The numerical procedures mirrors the MPS-SMA of Appendix~\ref{app:sma}.

\begin{enumerate}[leftmargin=*]
\item \textbf{Exact block elements.} On an open window of $N_w=16$ cells,
build the two-wall product states $\ket{i,d}$, $0\le d\le D$, centred on
the window, and evaluate every Hamiltonian and overlap element by exact
dense contraction,
\begin{equation}
\begin{split}
  M[\Delta,d,d']&=\bra{i,d}H\ket{i{+}\Delta,d'},\\
  \mathcal O[\Delta,d,d']&=\braket{i,d}{i{+}\Delta,d'},
\end{split}
\label{eq:blocks}
\end{equation}
which depend only on $(\Delta,d,d')$ by translation invariance in the
window bulk. The elements decay geometrically with the wall shift [as the
overlap $(-\tfrac12)^{|\Delta|}$, Eq.~\eqref{eq:overlap}] and are kept for
$|\Delta|\le4$, with both walls restricted to the bulk.

\item \textbf{Momentum assembly.} The centre-of-mass Fourier transform
$\ket{q,d}=L^{-1/2}\sum_i e^{iq(i+d/2)}\ket{i,d}$ turns the blocks into
the $(D{+}1)\times(D{+}1)$ matrices
\begin{equation}
  H(q)_{dd'}=\sum_{\Delta}e^{iq\Delta}\,e^{iq(d'-d)/2}\,M[\Delta,d,d'],
  \label{eq:Hq-assembly}
\end{equation}
and identically $\mathcal O(q)$ from $\mathcal O[\Delta,d,d']$. The
half-phase $e^{iq(d'-d)/2}$ is a diagonal gauge that leaves the spectrum unchanged but renders both matrices Hermitian.

\item \textbf{Canonical orthogonalization.} The $(-\tfrac12)^{|d|}$
overlaps make the two-wall states nearly linearly dependent, so
$\mathcal O(q)$ is near-singular and a Cholesky-based generalized solver
fails. We therefore diagonalize the metric,
$\mathcal O(q)=U\,\mathrm{diag}(s)\,U^\dagger$, discard the directions with $s_i\le\tau\max_j(s_j)$ ($\tau=10^{-8}$; typically one direction at $D=6$), build $X=U\,\mathrm{diag}(s)^{-1/2}$ on the kept directions [so that $X^\dagger\mathcal O(q)X=\mathbbm 1$], solve the standard Hermitian problem for $X^\dagger H(q)X$, and map the eigenvectors back as $\psi_n=X\phi_n$, which are $\mathcal O$-normalized.

\item \textbf{Source weights} The basis is
non-orthogonal, the overlap with the source is taken in the metric,
$w_n=|(\mathcal O(q)\psi_n)_0|^2$ [the weight of Eq.~\eqref{eq:2kink-sf}]; the total weight at each $q$, $\sum_n w_n=\mathcal O(q)_{00}$ up to the discarded null directions, is conserved and merely redistributed over the bands.

\item \textbf{Energy reference.} Energies are measured from the window VBS energy $E_0$. At $r=1$ the VBS is the true ground state and the spectrum is used as is. For $r\neq1$ the bands are shifted rigidly so that the $D=6$ continuum floor coincides with the extrapolated gap $\Delta_{\rm ED}$ of Eq.~\eqref{eq:anchor}, the same anchor used for the closed-form dispersions.
\end{enumerate}

The truncation yields $D{+}1$ bands that densify into the continuum as
$D\to\infty$. The $D=0$ term is the coincident pair, i.e.\ exactly the
triplon of Appendix~\ref{app:bondop}. Increasing $D$ lets the pair
delocalize and redistributes the weight downward from $\omega=\JAB$ toward the two-spinon gap.

% =====================================================================
\bibliographystyle{apsrev4-2}
\bibliography{bibliography}

\end{document}